\documentclass[sn-standardnature,iicol]{sn-jnl}

\usepackage[linesnumbered,algo2e]{algorithm2e}
\usepackage{pbox}

\jyear{2021}%

\theoremstyle{thmstyleone}%
\theoremstyle{thmstyletwo}%

\theoremstyle{thmstylethree}%

\usepackage{bm}

\begin{document}

\title[Cloud K-SVD for Image Denoising]{Cloud K-SVD for Image Denoising}

\author*[1,2]{\fnm{Christian Marius} \sur{Lillelund \url{https://orcid.org/0000-0002-2520-3774}}}\email{cl@ece.au.dk}
\equalcont{These authors contributed equally to this work.}

\author[2,3]{\fnm{Henrik Bagger} \sur{Jensen}}\email{201304157@post.au.dk}
\equalcont{These authors contributed equally to this work.}

\author[1,2]{\fnm{Christian Fischer} \sur{Pedersen \url{https://orcid.org/0000-0002-1266-5857}}}\email{cfp@ece.au.dk}

\affil*[1]{\orgdiv{Department of Electrical and Computer Engineering}, \orgname{Aarhus University}, \orgaddress{\street{Finlandsgade 22}, \postcode{DK-8200}, \city{Aarhus N}, \country{Denmark}}}


\abstract{Cloud K-SVD is a dictionary learning algorithm that can train at multiple nodes and hereby produce a mutual dictionary to represent low-dimensional geometric structures in image data. We present a novel application of the algorithm as we use it to recover both noiseless and noisy images from overlapping patches. We implement a node network in Kubernetes using Docker containers to facilitate Cloud K-SVD. Results show that Cloud K-SVD can recover images approximately and remove quantifiable amounts of noise from benchmark gray-scaled images without sacrificing accuracy in recovery; we achieve an SSIM index of 0.88, 0.91 and 0.95 between clean and recovered images for noise levels ($\mu$ = 0, $\sigma^{2}$ = 0.01, 0.005, 0.001), respectively, which is similar to SOTA in the field. Cloud K-SVD is evidently able to learn a mutual dictionary across multiple nodes and remove AWGN from images. The mutual dictionary can be used to recover a specific image at any of the nodes in the network.}


\maketitle

\section{Introduction}

\begin{figure}[!htb]
	\centering
	\includegraphics[width=1\linewidth]{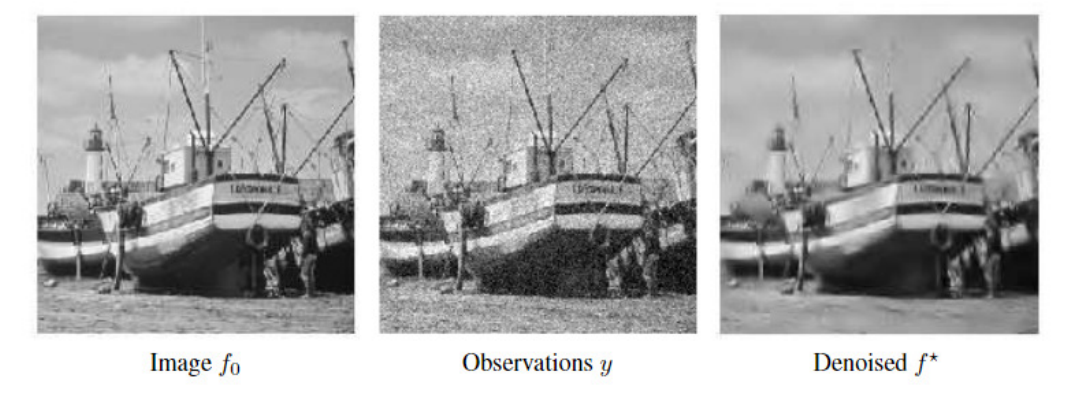}
	\caption{Left: Original. Middle: Noisy. Right: Denoised. Source: \cite{Mallat2009}}
	\label{fig:denoisingpicture}
\end{figure}

The Big Data era, in which we live in, generates large volumes of valuable data from a broad range of data sources \cite{Taylor-Sakyi2016}. Data gathered exclusively by social networks translates to millions of images, videos and messages alone. Almost 500TB social data is produced every day \cite{Li2016}. With this surge of data, the scientific community has had a growing interesting in sparse approximation of signals \cite{Elad2010} for applications such as compression, denoising, restoration and classification \cite{Elad2010, Elad2011, Aharon2006, Mairal2012, Theodoridis2015}. A central problem in sparse approximation is the design of a dictionary that contains fundamental signal atoms, where signals can be described as typically linear combinations of these atoms. For this task, we can either use predefined and prefixed dictionaries such as overcomplete wavelet or Gabor dictionaries \cite{Mallat2009}, or make our own by adapting a random dictionary to a set of training signals, a practice referred to as dictionary learning \cite{Elad2010, Elad2011, Aharon2006, Mairal2012, Theodoridis2015, Tosic2011}, for example the acclaimed K-SVD algorithm from 2006 by Aharon, Elad and Bruckstein \cite{Aharon2006}. K-SVD has since been widely adopted, for example to recover corrupted imaging of clouds and shadows using temporal correlations \cite{Li2014} or to improve quality in images of feathers by utlizing K-SVD to train a global dictionary that can remove unwanted noise \cite{Yan2017}. In 2016, Raja and Bajwa \cite{Raja2016} published Cloud K-SVD, an extension to traditional K-SVD that offers several benefits when data is bulky or governed by privacy concerns: Cloud K-SVD can train at multiple local or distributed nodes and hereby produce a mutual dictionary to represent low-dimensional geometric structures in all the data combined. This is accomplished without any exchange of full or partial data sets between nodes. This is a desirable trait when data is governed by privacy concerns or comes in large volumes. In fact, since the processing of the data is performed locally and the nodes exchange (intermediate) training results in the shape of residual errors instead of raw data, the need for sensitive information exchange is bypassed \cite{Clifton2002}. The nodes use averaging consensus to reach an agreement on the importance of each dictionary atom, thus they collaboratively exchange and average atoms from their local dictionary until they are in concordance of a global one. We present a novel application of Cloud K-SVD, as we implement it in a real node network and use it to remove Additive White Gaussian Noise (AWGN) (figure \ref{fig:denoisingpicture}). Each node is responsible for denoising a certain part of the image, but collaborates with fellow nodes to reach a consensus of a global dictionary in the process. This produces a sparse, noise reduced image and a global dictionary, which encompasses information present at all nodes. In practice, we implement the algorithm using Docker containers and run experiments in Kubernetes. We compare our results in terms of PSNR and SSIM of recovered images with the Pruned Non-Local Means (PNLM) algorithm by Ghosh, Chaudhury and Mandal presented in \cite{Ghosh2017}. Source code is available on GitHub \cite{Lillelund2021}.

Section \ref{sec:theory} presents the theoretical background to the methods in use. Section \ref{sec:setup} describes our practical setup and the hardware used. Section \ref{sec:exp_and_results} contains all experiments and their results. Lastly section \ref{sec:conclusion} provides a summary of the work and some final remarks.

\section{Theory on signal processing}
\label{sec:theory}

This section describes the main concepts: Sparse approximation, dictionary learning, consensus averaging and the Cloud K-SVD algorithm.

\subsection{Sparse approximation}

In sparse approximation, the objective is to provide the simplest possible approximation of a signal $\bm{x}$ as a linear combination of fewest possible columns, i.e. atoms, from a dictionary $\bm{D} \in \mathbb{R}^{M \times N}$~\cite{Elad2010}. This dictionary is typically overcomplete ($M<N$), which leads to an underdetermined linear system $\bm{y} = \bm{Dx}$, where $\bm{y} \in \mathbb{R}^M$ is the observed signal and $\bm{x} \in \mathbb{R}^N$ is unobserved signal vector. Since $\bm{D}$ is underdetermined, there are infinitely many solutions; among these, the objective is to find the sparsest (fewest nonzeros) possible representation $\bm{x}$ satisfying $\bm{y} = \bm{Dx}$, i.e. $\min ||\bm{x}||_0 ~s.t.~ \bm{y} = \bm{Dx}$, where $\lVert \cdot \rVert_{0}$ denotes the $\ell_0$ pseudo/counting norm. Often the observed signal $\bm{y}$ is noisy, which introduces a noise budget and a relaxation of the equality constraint in the minimization problem: $\min ||\bm{x}||_0 ~s.t.~ || \bm{y} - \bm{Dx}||_2^2 \leq \varepsilon^2$. This is a NP-Hard problem; commonly the norms are changed from $\ell_0$ to $\ell_1$ and the solution is approximated by a pursuit algorithm, e.g. Orthogonal Matching Pursuit \cite{Mallat1993}. An equivalent minimization problem can be formulated in matrix notation by:

\begin{equation}
	\min_{\bm{X} \in \mathbb{R}^{N \times Q}} || \bm{Y} - \bm{DX} ||_{2} \leq \epsilon \text{~~s.t.~~} \forall i: || \bm{x}_{i} ||_{0}  \leq K
\end{equation}
\noindent{where $\bm{E, Y} \in \mathbb{R}^{M \times Q}$, $\bm{X} \in \mathbb{R}^{N \times Q}$ and $K$ is the number of nonzero elements in the columns of $\bm{X}$.}

\subsection{Dictionary learning}

We solve the linear inverse problem $\bm{Y} = \bm{D} \bm{X}$ for $\bm{X}$ to recover an approximation of signal $\bm{X}$ from $\bm{Y}$ sampled measurements. The dictionary $\bm{D}$ guides our recovery process in using the best nonzero transform coefficients because it contains a database of geometric features in the data of interest. Dictionary learning is a framework to record these features from sampled measurements \cite{DumitrescuBogdanIrofti2018}. This assumes samples, e.g. image patches, can be sparsely approximated by a linear combination of a few columns from a suitable overcomplete basis. Let $\bm{Y} \in \mathbb{R}^{M \times Q}$ be a matrix of vectorized image patches, then each patch, $\bm{y}_{i} \in \mathbb{R}^{M}$, has a sparse representation in a dictionary, $\bm{D} \in \mathbb{R}^{M \times N}$, if $\bm{Y} \approx \bm{D} \bm{X}$ and the sparse signal matrix $\bm{X} \in \mathbb{R}^{N \times Q}$, whose columns have at most $K$ nonzero elements \cite{DumitrescuBogdanIrofti2018}. The dictionary $\bm{D}$ is then a dimensionality reduction step and can be either predefined or learned from data \cite{Aharon2006, Elad2011, Mairal2010}.

\subsection{Averaging consensus}

Averaging consensus \cite{Chen2011} is a method to let nodes collaboratively share their state, for example the current residual vector in a dictionary learning scenario, then by averaging steer the pool of nodes towards a mutual averaged consensus. Consider a multi-hop network of a group of $H$ nodes modeled as a directed graph, $\mathcal{G} = (\mathcal{V},\varepsilon)$, where $\mathcal{V} = {1,2, \dots,H}$ represents the $H$ nodes and the edge set $\varepsilon = {(i,j) : i,j \in \mathcal{V}, i \neq j}$ consists of vertex pairs. A vertex pair $(i,j)$ represents the directed edge from $i$ to $j$, i.e. node $i$ can send traffic to node $j$. The averaging protocol assumes each node $i$ holds some value ${z}_{i} \in \mathbb{R}$ and computes the average ${\bar{z}} = \sum_{i=1}^{H} {z}_{i} / H$ using linear distributed iterations. Each node have a local state variable ${x}_{i}(t=0) = {z}_{i}$ and iteratively updates the value with a weighed average of its neighbors' state variables. The state variable ${x}_{i}(t)$ converges to ${\bar{z}}$ under certain conditions \cite{Chen2011}. After receiving their neighbors' state, nodes update their state variables as such:

\begin{equation}
	\begin{split}
		{x}_{i}(t + 1) = {x}_{i}(t) + \\ \sum_{j=1,j \neq i}^{H} \bm{W}_{ij}(t) \Big({x}_{j}(t) - {x}_{i}(t)\Big)
	\end{split}
\end{equation}
\noindent{where $\bm{W}(t) \in \mathbb{R}^{H \times H}$ is a weight matrix defined as $\bm{W}(t) := \bm{I} - \varepsilon \bm{L}(t)$, where $\bm{I}$ is the identity matrix and $\bm{L} = [l_{ij}]$ is the graph Laplacian matrix of the network \cite{Chen2011}}.

\subsection{Cloud K-SVD}
\label{subsec:cloud_ksvd}

\SetKwInOut{Initialize}{Initialize}
\begin{algorithm}[ht]
	\KwIn{Local data $\bm{Y}_{1}, \bm{Y}_{2}, \dots, \bm{Y}_{H}$, parameters $N$ and $K$ and doubly-stochastic matrix $\bm{W}$}
	\BlankLine
	\Initialize{Generate $\bm{d}^{ref} \in \mathbb{R}^{M}$ and $\bm{D}^{init} \in \mathbb{R}^{M \times Q}$ randomly, set $t_{d} \leftarrow 0$ and $\bm{\hat{D}}_{i}^{(t_{d})}$ $\leftarrow$ $\bm{\hat{D}}_{}^{init}$, $i = 1, 2, \dots, H$.}
	\While{stopping criteria not satisfied}{
		$t_{d} \leftarrow t_{d} + 1.$
		\BlankLine
		The $i^{th}$ site solves $\forall q, \bm{\hat{x}}_{i,q}^{(t_{d})} \leftarrow \arg \min\limits_{\bm{x} \in \mathbb{R}^{K}} \parallel \bm{y}_{i,q} - \bm{\hat{D}}_{i}^{(t_{d}-1)} \bm{x} \parallel_{2}^{2}$ $\;$ s.t. $\;$ $\parallel \bm{x} \parallel_{0} \leq K$ using SOMP
		\BlankLine
		\For(\tcp*[f]{Dictionary Update}){$n = 1$ \KwTo $N$}{
			$\bm{\hat{E}}_{i,n,R}^{(t_{d})} \leftarrow \bm{Y}_{i} \bm{\hat{\Omega}}_{i,n} - \sum_{j=1}^{n-1} \bm{\hat{d}}_{i,j}^{(t_{d})} \bm{\hat{x}}_{i,j,T}^{(t_{d})} \bm{\hat{\Omega}}_{i,n}^{(t_{d})} - \sum_{j=n+1}^{N} \bm{\hat{d}}_{i,j}^{(t_{d}-1)} \bm{\hat{x}}_{i,j,T}^{(t_{d})} \bm{\hat{\Omega}}_{i,n}^{(t_{d})}$
			\BlankLine
			$\bm{\hat{M}}_{i} \leftarrow \bm{\hat{E}}_{i,n,R}^{(t_{d})} \bm{\hat{E}}_{i,n,R}^{{(t_{d})}^{T}}$
			\BlankLine
			Generate $\bm{q}^{init}$, $t_{p} \leftarrow 0$ and $\bm{\hat{q}}_{i}^{(t_{p})} \leftarrow \bm{q}^{init}$
			\BlankLine
			\While(\tcp*[f]{Power Method}){stopping rule}{
				$t_{p} \leftarrow t_{p} + 1$
				\BlankLine
				Set $t_{c} \leftarrow 0$ and $\bm{z}_{i}^{(t_{c})} \leftarrow \bm{\hat{M}}_{i} \bm{\hat{q}}_{i}^{t_{p}-1}$
				\BlankLine
				\While(\tcp*[f]{Consensus}){stopping rule}{
					$t_{c} \leftarrow t_{c} + 1$
					\BlankLine
					$\bm{z}_{i}^{(t_{c})} \leftarrow \sum_{j \in \mathcal{N}_{i}} \bm{w}_{i,j} \bm{z}_{i}^{(t_{c}-1)}$ \\$\iff$ the $j^{th}$ site has $\bm{z}_{j}^{(t_{c}-1)}$
					\BlankLine
				}
				$\bm{\hat{q}}_{i}^{(t_{p})} \leftarrow \bm{\hat{v}}_{i}^{(t_{p})} / \parallel \bm{\hat{v}}_{i}^{(t_{p})} \parallel_{2}$\;
			}
			$\bm{\hat{d}}_{i,n}^{(t)} \leftarrow \text{sgn} \Big( \langle \bm{d}^{ref}, \bm{\hat{q}}_{i}^{(t_{p})} \rangle \Big) \bm{\hat{q}}_{i}^{(t_{p})}$
			\BlankLine
			$\bm{\hat{x}}_{i,n,R}^{(t)} \leftarrow \bm{\hat{d}}_{i,n}^{(t_{d})^{T}} \bm{\hat{E}}_{i,n,R}^{(t_{d})}$
			\BlankLine
		}
	}
	\KwResult{$\bm{\hat{D}}_{i}^{t_{d}}$ and $\bm{\hat{X}}_{i}^{t_{d}}$, $i = 1, 2, \dots, H$}
	\caption{The cloud K-SVD algorithm. Originally by \cite{Raja2016}. Modifications made by authors.}
	\label{alg:cloudksvd}
\end{algorithm}

We assume a distributed setting, where local data is held at different nodes. Each node is denoted by $H_{i}$ with local data $\bm{Y}_{i} \in \mathbb{R}^{M \times Q_{i}}$ for a total of $H$ nodes, so the total data amount is $Q = \sum_{i=1}^{H} Q_{i}$ and can be represented as one matrix $\bm{Y} \in \mathbb{R}^{M \times Q}$. Our algorithm is presented in \ref{alg:cloudksvd} based on \cite{Raja2016}. Modifications made by authors include use of Simultaneous Orthogonal Matching Pursuit (SOMP) over regular OMP in \cite{Raja2016} (line 3) and a condition when computing the residual vector $\bm{z}_{i}^{(t_{c})}$ that the neighboring $j^{th}$ site must be reachable and hold $\bm{z}_{j}^{(t_{c}-1)}$ (line 13), otherwise its residual is not included when updating the $\bm{\hat{d}}_{i,n}^{(t)}$ dictionary element. We will now review the algorithm. First step is to compute the local signal matrix $\bm{\hat{X}}_{i}^{(t_{d})}$ using any sparse approximation algorithm. This step is as follows:

\begin{equation}
	\begin{split}
		\forall q, \bm{\hat{x}}_{i,q}^{(t_{d})} = arg \min_{\bm{x} \in \mathbb{R}^{N}} \parallel \bm{y}_{i,q} - \bm{\hat{D}}_{i}^{(t_{d}-1)} \bm{x} \parallel_{2}^{2} \\ \text{subject to} \parallel \bm{x} \parallel_{0} \leq K
	\end{split}
\end{equation}
\noindent{where $\bm{y}_{i,q}$ and $\bm{\hat{x}}_{i,q}^{(t_{d})}$ denote the ${q^{th}}$ sample and the signal vector at node $i$, respectively.} \\

Computing the sparse coefficients locally is acceptable at each iteration as long as the dictionary atoms in $\bm{\hat{D}}_{i}^{(t_{d}-1)}$ remain close to each other \cite{Raja2016}. Next is the dictionary update step. For this, Cloud K-SVD uses a distributed power iterations model to find the dominant eigenvector denoted $\bm{q}$ of a square, positive-semidefinite residual matrix $\bm{M}$, which is defined to have no eigenvalues less than or equal to zero. The power method is denoted as $\bm{\hat{q}}^{(t_{d})} = \bm{M} \bm{\hat{q}}^{(t_{d}-1)}$, where $\bm{\hat{q}}^{(t_{d})}$ is the estimate of the dominant eigenvector and approaches $\bm{q}$ for every iteration $(t_{d} \rightarrow \infty)$. We denote the initial value of $\bm{q}$ as $\bm{q}^{init}$ and may be a non-zero vector. All nodes begin with a starting vector $\bm{q}^{init}$ and matrix $\bm{M}_{i}$. At each iteration of the power method, the nodes reach a average consensus on an estimate of $\bm{q}$ for the matrix $\bm{M}$, defined as $\bm{M} = \sum_{i=1}^{H} \bm{M}_{i}$, where $H$ represents the number of nodes. If we apply the power method to $\bm{M}$, we observe:

\begin{equation}
	\begin{split}
	\bm{\hat{q}}^{(t_{d})} = \bm{M} \bm{\hat{q}}^{(t_{d}-1)} = \Big(\sum_{i=1}^{H}\bm{M}_{i}\Big)\bm{\hat{q}}^{(t_{d}-1)} \\ = \sum_{i=1}^{H}\bm{M}_{i}\bm{\hat{q}}_{i}^{t_{d}-1}
	\end{split}
\end{equation}

Each node maintains a $\bm{\hat{q}}^{(t_{d}-1)}$ that holds the previous error. Consensus averaging is then used to find the last summation of the above equation. Since the output $\bm{\hat{q}}^{(t)}$ requires both consensus averaging and the power method step, the local estimate at each node is affected by an error from both algorithms. If we perform enough consensus iterations $t_{c}$ and power iterations $t_{p}$, each node's estimate of the dominant eigenvector will converge to the true value for the residual $\bm{M}$ \cite{Tropp2006}. To compute an estimate of a local dictionary, given by $\bm{\hat{D}}_{i}^{t_{d}} \forall i = 1, 2, \dots, H$, to form a global dictionary estimate given by $\bm{\hat{D}}^{t_{d}}$ at iteration $t_{d}$, we update atoms $\bm{d}_{i,n}$ at each node by collaboratively finding the total error to be minimized. This requires us to find local error matrix called $\bm{\hat{E}}_{i,n,R}^{(t_{d})}$. To do so, we need to find the indices used by the dictionary at each node, denoted $\bm{\omega}_{n}$. In this case, $\bm{\omega}_{i,n}^{t_{d}}$ refers to the indices used by the dictionary atom $\bm{D}_{i,n}$ at iteration $t_{d}$:

\begin{equation}
	\bm{\omega}_{i,n}^{(t_{d})} = \Big\{q \parallel 1 \leq q \leq Q_{i}, \bm{x}_{i,n,T}^{(t_{d})}(q) \neq 0\Big\} \forall i
\end{equation}

\noindent{where $\bm{x}_{i,n,T}^{(t_{d})}(q)$ denotes the $q^{th}$ element of $\bm{x}_{i,n,T}^{(t_{d})}$.}

We define $\bm{\Omega}_{i,n}^{(t_{d})} = Q \times \bm{\omega}_{i,n}^{(t_{d})}$ as a binary matrix with ones in $(\bm{\omega}_{i,n}^{(t_{d})} (q), q)$ to store $\bm{\omega}_{i,n}^{(t_{d})}$. Thus $\bm{\hat{E}}_{i,n,R}^{(t_{d})} = \bm{\hat{E}}_{i,n}^{(t)} \bm{\Omega}_{i,n}^{(t_{d})}$. $\bm{\hat{E}}_{i,n}^{(t_{d})}$ is given with this equation:

\begin{equation}
	\begin{split}
		\bm{\hat{E}}_{i,n}^{(t_{d})} = \bm{Y}_{i} - \sum_{j=1}^{n-1} \bm{\hat{d}}_{i,j}^{(t_{d})} \bm{\hat{x}}_{i,j,T}^{t_{d}} \\ - \sum_{j=n+1}^{N} \bm{\hat{d}}_{i,j}^{(t_{d}-1)} \bm{\hat{x}}_{i,j,T}^{t_{d}-1}
	\end{split}
\end{equation}

\noindent{where $j$ is the dictionary atom for $1, 2, \dots, N$, except for $j=i$.}

Here we update the atoms in numerical order. Updated atoms are in iteration $t_{d}$ and not updated atoms are in $t_{d}-1$. Next step is to calculate the SVD as in K-SVD of $\bm{\hat{E}}_{n,R}^{(t_{d})} = \sum_{i=1}^{N} \bm{\hat{E}}_{i,n,R}^{(t_{d})}$ for all nodes to find $\bm{U}_{1,n}$ and $\bm{V}_{1,n}$. We make the following updates: $\Big(\bm{\hat{d}}_{i,n}, \bm{\hat{x}}_{i,n,R}) = (\bm{U}_{1,n}, \bm{\Delta}_{(1,1)} \bm{V}_{1,n}) \forall i$
By definition we know:

\begin{equation}
	\begin{split}
		\bm{\hat{x}}_{i,n,R} = \bm{\hat{d}}_{i,n} \bm{\hat{E}}_{i,n,R}^{(t_{d})} = \bm{U}_{1,n} \bm{\hat{E}}_{i,n,R}^{(t_{d})} \\ = \bm{\Delta}_{(1,1)} \bm{V}_{1,n} \forall i
	\end{split}
\end{equation} 

To find an approximation of $\bm{\hat{E}}_{i,n,R}^{(t_{d})}$, we use a distributed power method to find its dominant eigenvector, i.e. $\bm{\hat{E}}_{n,R}^{(t_{d})}$ need be a square and positive semi-definite matrix. This can be archived by multiplying the matrix by its transpose, $\bm{\hat{E}}_{n,R}^{(t_{d})} \bm{\hat{E}}_{n,R}^{(t_{d})^{T}}$, and applying the power method to the resulting matrix defined as:

\begin{equation}
	\bm{\hat{M}}^{(t_{d})} = \sum_{n=1}^{H} \bm{\hat{M}}_{n}^{(t_{d})} = \sum_{n=1}^{H} \bm{\hat{E}}_{n,R}^{(t_{d})} \bm{\hat{E}}_{n,R}^{(t_{d})^{T}}
\end{equation}

We then find the dominant eigenvector of $\bm{\hat{M}}_{n}^{(t_{d})}$. By definition, it is the $U_{1}$ of $\bm{\hat{E}}_{n,R}^{(t_{d})}$, since for any real matrix $\bm{A}$, the left-singular vectors of $\bm{A}$ are the eigenvectors $\bm{A} \bm{A}^{T}$. With this in mind, $\bm{\hat{d}}_{i,n}$ is the result of our distributed power method. We locally compute $\bm{\hat{d}}_{i,n}^{(t_{d})^{T}} \bm{\hat{E}}_{i,n,R}^{(t_{d})}$ to set $\bm{\hat{x}}_{i,n,R}$ and lastly set $\bm{\hat{x}}_{i,n} = \bm{\hat{x}}_{i,n,R} \bm{\Omega}_{i,n}^{(t_{d})^{T}}$ to place the zeros into the row of coefficients. Now the dictionary update for one atom is done and when all atoms have been updated, the dictionary update step is complete. The loop repeats for each dictionary learning iteration $(t_{d})$.

\section{Node network setup}
\label{sec:setup}

We build a Kubernetes network of four 4 Raspberry Pi 2 Model B nodes to perform our practical experiments on. In Kubernetes, a deployment controller is configured for the $P$ worker pods. Each pod can receive, compute and send data as a self-contained and autonomous unit. We scale the number of pods manually for a deployment to demonstrate efficiency of the Cloud K-SVD algorithm when more pods are added to the working pool and input data is further distributed. All pods share the same networking media, so potential network delays are included in the results.

\begin{figure}[!htb]
	\centering
	\fbox{\includegraphics[width=1\linewidth]{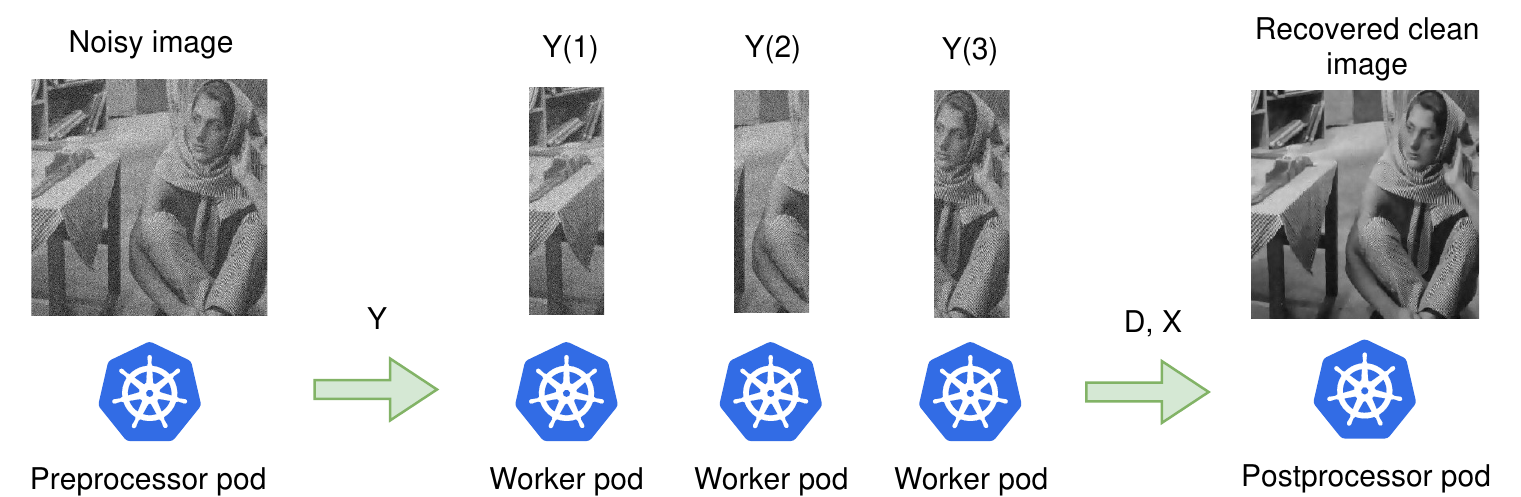}}
	\caption{The Kubernetes orchestration: E.g., three worker pod replicas running Cloud K-SVD exist. The preprocessor splits the image data in three equally-sized parts. The worker pods process their respective part of the image and the postprocessor puts it all back together.}
	\label{fig:denoisingKubernetes}
\end{figure}

Pod $P_{i}$ receives $\bm{Y}_{i} \in \mathbb{R}^{M \times Q_{i}}$ data signals and perform dictionary learning collaboratively by first calculating the signal vectors $\bm{X}_{i}$ via sparse approximation and then estimate the residual error via consensus iterations. Consensus averaging is used to lower the residual error and properly estimate the signal vectors by a local dictionary $\bm{D}_{i}$ that accommodates data at pod $i$. All experiments are done with one preprocessing and one postprocessing pod (figure \ref{fig:denoisingKubernetes}). These pods are not active in the experiments and are used only to split and aggregate data.

\begin{figure}[!htb]
	\centering
	\includegraphics[width=1\linewidth]{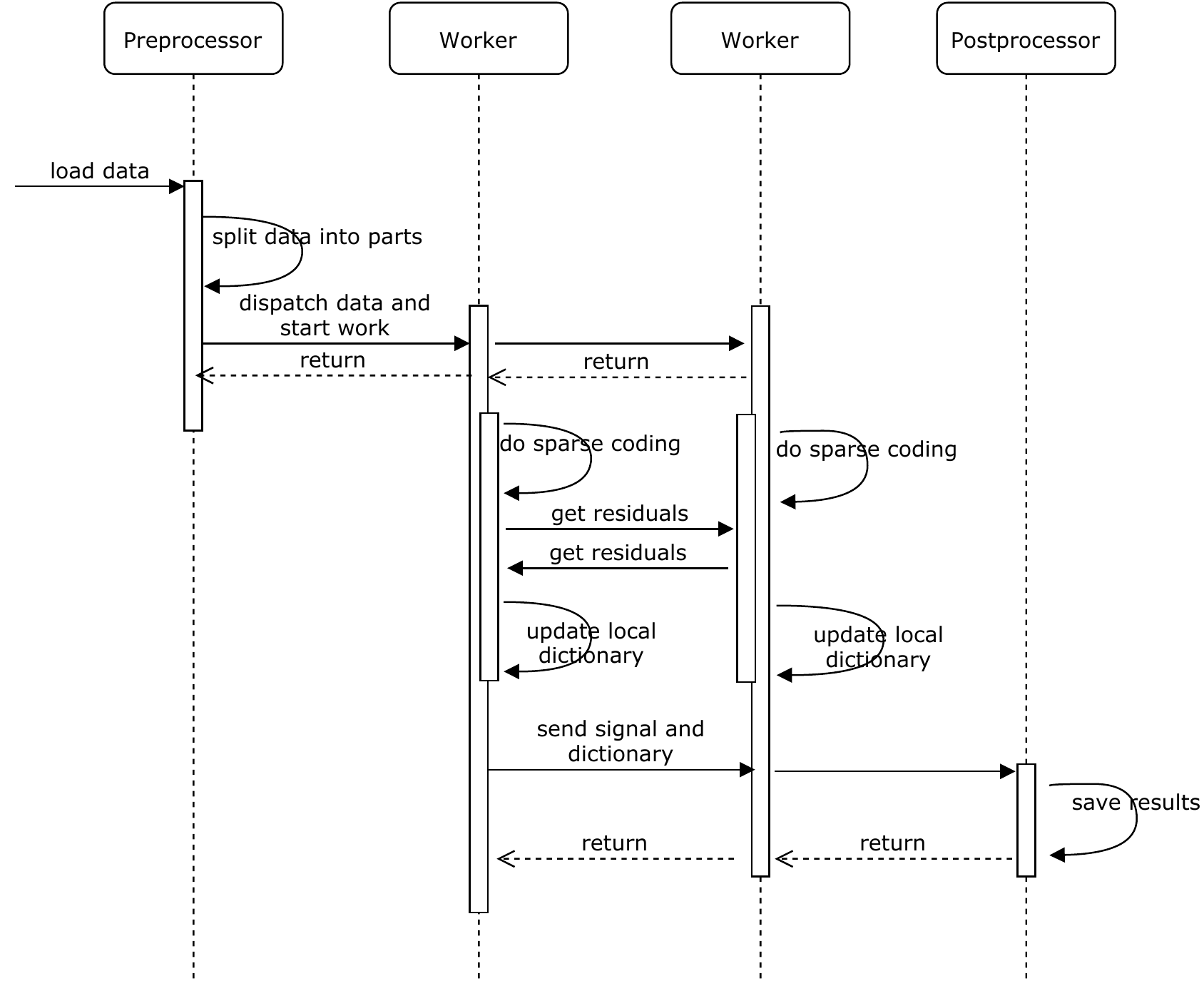}
	\caption{In this example, data is loaded onto two worker pods. Cloud K-SVD is then started and at the end a signal matrix and a local dictionary is produced, which the postprocessor pod collects.}
	\label{fig:sequenceDiagram}
\end{figure}

Figure \ref{fig:sequenceDiagram} shows a small sequence diagram of the interaction between the preprocessor, worker and postprocessor pod. It shows all steps in our implementation and the cloud K-SVD protocol at a high level. The drawing has been simplified a bit, hence data is sent from the preprocessor to individual workers directly, not through each other as it may seem. This is also the case when data is saved in the end.

\section{Experiments and Results}
\label{sec:exp_and_results}

Our experiments looks to demonstrate Cloud K-SVD's ability to learn a new dictionary from completely uncharted clean and noisy image patches and recover these as a sparse approximation. Recovery accuracy is the chosen performance metric, which is a function of the number of iterations, $t_{d}$, length of signals (patches as column vectors), $M$, number of signals (patches), $Q$, and the atom count, $N$, in the dictionaries. We consider a case, where patches have been contaminated with AWGN commonly found in natural images.

\subsection{Variants of K-SVD}

We consider two variants of K-SVD:

\begin{itemize}
\item Local K-SVD: Pod $P_{i}$ receive $\bm{Y}_{i} \in \mathbb{R}^{M \times Q_{i}}$ data signals and perform the dictionary learning task locally without any collaborative consensus iterations among peers. This model distributes the data signals, but pod $P_{i}$ will not become familiar with pod $P_{j}$'s data due to no consensus iterations. This model mimics the traditional K-SVD scheme, but distributes the data among multiple working pods. 
\item Cloud K-SVD: Pod $P_{i}$ receive $\bm{Y}_{i} \in \mathbb{R}^{M \times Q_{i}}$ data signals and perform dictionary learning collaboratively by first calculating the signal vector $\bm{X}_{i}$ via sparse approximation and then estimate the residual error via consensus iterations. Here, all worker pods $P$ work together by either averaging or corrective consensus to lower the residual error and properly estimate the signal vectors by a local dictionary $\bm{D}_{i}$ that accommodates data at node $j$.
\end{itemize}

\subsection{Parameter Settings}
\label{subsec:parametersettings}

The full size of the training data is $M \times Q$, where $M$ is the length of the data signals in $\bm{Y} \in \mathbb{R}^{M \times Q}$ and $Q$ is the number of signal column vectors in $\bm{Y}$ and $\bm{X}$. Image patch data (columns) will be used to train the distributed dictionaries; thus, the size of an input patch is $M$. The resulting data vectors are gathered in a single matrix $\bm{Y} \in \mathbb{R}^{M \times Q}$. We then add AWGN to all training data in $\bm{Y}$ at the preprocessor pod. Data is then divided into equally-sized parts and distributed to $P$ worker pods (figure \ref{fig:denoisingKubernetes}). Worker pods are now responsible for making a sparse approximation that reduces the amount of AWGN in the resulting $\bm{\hat{Y}}$ data matrix. For each experiment, a unique random (i.i.d. uniformly distributed entries with $\ell_{2}$ normalized columns in the range $[0;1]$) dictionary $\bm{D}^{M \times N}$ is generated and distributed to each worker pod. Then, all worker pods in the network are in charge of their own local dictionary, which they will evolve as the training process commences. We use our own implementation of the Simultaneous Orthogonal Matching Pursuit (SOMP) algorithm \cite{Tropp2006} and assign the weight scalar $w$ values in the range $[0;1]$. The following list of parameter settings hold for all experiments:

\begin{itemize}\setlength\itemsep{0em}
	\item Cloud K-SVD performs three types of nested iterations:
	\begin{itemize}
		\item $t_{d}$: A number of cloud iterations that compute the sparse signal vector via SOMP and update the corresponding atoms in the dictionary update step. \textbf{Setting:} $t_{d} = 10$.
		\item $t_{p}$: A number of power iterations for each atom $N$. The number is a trade-off between execution time and degree of consensus. \textbf{Setting:} $t_{p} = 3$.
		\item $t_{c}$: A number of consensus iterations that fetches the residual error from neighboring pods by averaging consensus. The number is a trade-off between execution time and degree of consensus. \textbf{Setting:} $t_{c} = 5$.
	\end{itemize}
	\item $N$: The number of atom column vectors in the dictionary $\bm{D} \in \mathbb{R}^{M \times N}$ and the length of the signal row vectors $\bm{X} \in \mathbb{R}^{N \times Q}$. As we evaluate overcomplete dictionaries in the experiments, it follows that $N \gg M$. The initial dictionary at pods is initialized with a fixed number atoms $N$ and dimension $M$ with randomly generated data. \textbf{Setting:} $N = 50, 100, 200$.
	\item $K$: The number of nonzero coefficients in the signal vectors of $\bm{X}$. \textbf{Setting:} $K = 3, 5, 7$.
	\item Pod weight $w$: A scalar multiplied by the difference between residual vectors $q_{i}$ and $q_{j}$ in the consensus step. It defines how much emphasis to put on vectors $q_{j}$ at node $i$ in range $[0,1]$. \textbf{Setting:} $w = 1$.  
	\item $P$: No. of worker pod replicas in Kubernetes that run Cloud K-SVD in a single experiment. \textbf{Setting:} $P=4$.
	\item $PS$: Patch size is the resolution of the patches extracted from training images; we require that $N \gg
	M$. \textbf{Setting:}
	\begin{itemize}
		\item Patch $5 \times 5:~~M=25 \text{~and~} N=50$
		\item Patch $6 \times 6:~~M=36 \text{~and~} N=100$
		\item Patch $7 \times 7:~~M=49 \text{~and~} N=100$
		\item Patch $8 \times 8:~~M=64 \text{~and~} N=150$
		\item Patch $9 \times 9:~~M=81 \text{~and~} N=200$
	\end{itemize}
	
\end{itemize}

\subsection{Experiment: Dictionary consensus}
\label{exp:consensus}

\begin{figure}[!htb]
	\centering
	\includegraphics[width=1\linewidth,trim={90 8 140 30},clip]{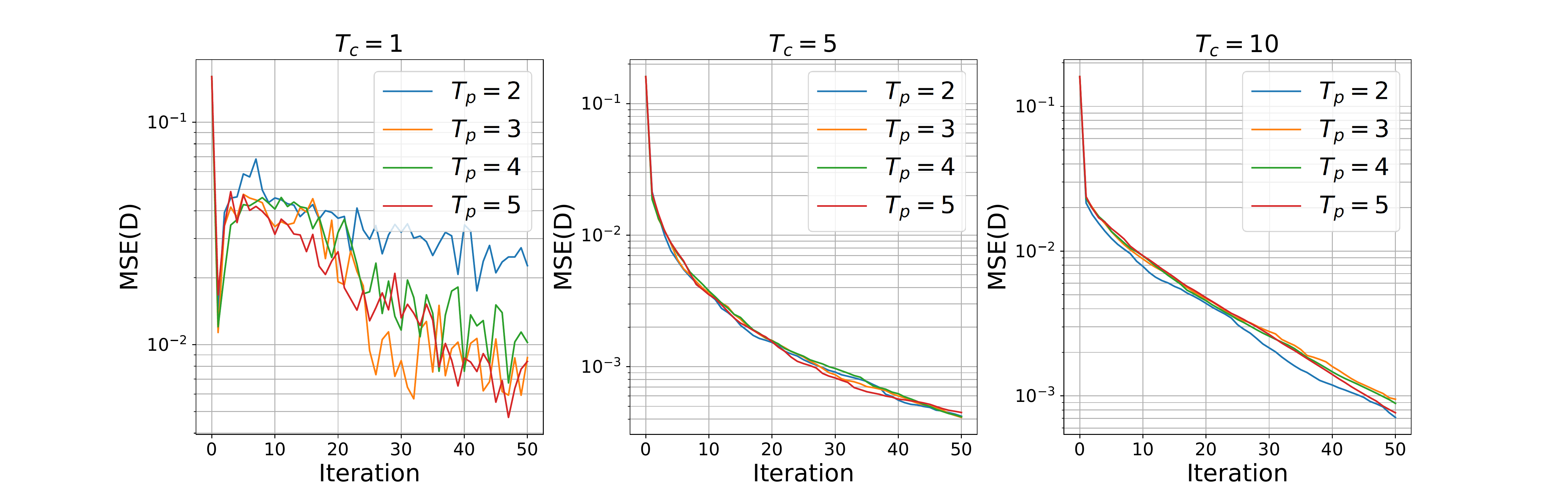}
	\caption{The average MSE across the pod dictionaries at different power iterations, $t_{p}$, and consensus iterations, $t_{c}$. We set cloud iterations, $t_{d} = 0, 1, \dots, 50$, number of atoms, $N=50$, length of the signals, $M=20$, number of signals, $Q=2000$, number of nonzero elements, $K=3$, and number of pods, $P=4$.}
	\label{fig:syn1}
\end{figure}

This experiment with synthetic data is to demonstrate the consensus element in Cloud K-SVD. Figure \ref{fig:syn1} shows the average MSE between dictionaries $\bm{D}_{i} \; \forall i = 1, 2, \dots$ as a function of $t_{d}$ iterations. Here, we fix the parameter $t_{c}$ for all experiments and notice the change in error values when we increase $t_{p}$. We see a lack of convergence for $t_{c}=1$ no matter the number of $t_{p}$, though when $t_{c}$ is higher, it produces a clear picture of the error between the dictionaries as they drop collectively towards zero on a logarithmic scale.

Table \ref{tab:syn1} shows a list of average execution times for $P=4$ pods using synthetic data tracked for the collaborative K-SVD step. The purpose is to show how consensus and power iterations impact the time it takes to complete the K-SVD step on average in the algorithm measured in seconds. The signal approximation step is not included here because it does not depend on the number of consensus or power iterations, thus remain unchanged throughout all experiments here. The trend is clear, however. The more consensus iterations we do per power iteration, the longer the algorithm takes to complete the K-SVD step.

\begin{table}[!htb]
	\begin{minipage}{1\linewidth}
		\caption{The average execution time of a K-SVD iteration with different $t_{p}$ and $t_{c}$.  $t_{d} = 50$, $M = 20$, $N = 50$, $Q = 2000$, $K = 3$, $P = 4$.}
		\label{tab:syn1}
		\begin{tabular}{lllll} \toprule
			{$t_{c}$} & {$t_{p}=2$} & {$t_{p}=3$} & {$t_{p}=4$} & {$t_{p}=5$} \\ \midrule
			1  & 4.5s & 6.6s & 8.3s & 9.8s \\
			2  & 18.4s  & 26.9s & 35.2s  & 43.1s \\
			3  & 34.5s  & 51.2s & 67.9s  & 84.2s \\ \bottomrule
		\end{tabular}
	\end{minipage}
\end{table}

\subsection{Experiment: Patch learning}

The second set of experiments evaluate Cloud K-SVD on clean images. The dataset consists of eight gray-scaled 8-bit images chosen with respect to their dimensionality and variety in contours. Table \ref{tab:imageInfo} shows properties such as the original resolution, the resolution when it has been downscaled by a factor $\alpha = 2$, image mean and variance.

\begin{table}[h]
	\begin{minipage}{1\linewidth}
		\caption{Test images for patch learning with their respective attributes. Downscaling by a decimation factor $\alpha$ creates blocks of $\alpha \times \alpha$ size and reduces elements in each block to a local mean.}
		\label{tab:imageInfo}
		\begin{tabular}{@{}llll@{}}
		\toprule
			Name & Original res & Res/$\alpha=2$ & Mean/variance \\
			\midrule
	Castle & $304 \times 200$ & $152 \times 100$ & 115/11.18  \\
	Lenna & $480 \times 512$ & $240 \times 256$ & 124/8.96  \\
	China & $427 \times 640$ & $214 \times 320$ & 114/26.95 \\
	Flower & $427 \times 640$ & $214 \times 320$ & 68$\;\:$/9.97\\
	Chelsea & $300 \times 451$ & $150 \times 226$ & 117/4.05\\
	Camera & $512 \times 512$ & $256 \times 256$  & 118/15.07\\
	Astronaut & $512 \times 512$ & $256 \times 256$ & 112/22.27 \\
	Face & $150 \times 150$ & $75 \;\: \times 75$ & 139/15.52 \\
			\botrule
		\end{tabular}
	\end{minipage}
\end{table}

For the sake of performance, we apply a decimation factor $\alpha$ in the range $[1-5]$ to some images before processing. $\alpha=1$ means no downscaling takes place, whereas $\alpha=2$ means the total number of pixels have been cut in half, column and row wise, compared to the original. For patch learning, we introduce a variable $PS$ which denotes the resolution of extracted patches, for example $5 \times 5$ or $8 \times 8$. For both cases, all possible overlapped patches of sizes $PS$ are extracted from each training image, for example $Q = (256 - 8 + 1)^{2} = 62001$ patches in total for a patch size of $PS = 8 \times 8$, resolution $256 \times 256$ and $M = 64$. The dimension of an input data sample is then $M$ (vertically stacked columns) for patch size $PS$ and the total amount of training data is then $T_{t} = M \times Q$. The resulting data vectors are gathered in a single matrix $Y \in \mathbb{R}^{M \times Q}$ before being split into equally-sized parts and distributed to $P$ pods for learning.

For each experiment, a starting dictionary $\bm{D}_{i}$ of size $M \times N$ is generated for each pod with random i.i.d. uniformly distributed entries with normalized $\ell_{2}$ columns in the range $[0,1]$. Each dictionary is distributed to pods $P_{i}$ before start. All worker pods are then in charge of their own local dictionary. Also we assign the weight scalar $w$ values in the range $[0,1]$.

\begin{itemize}
	\item The patch size is PS = 5 $\times$ 5, 6 $\times$ 6, 7 $\times$ 7, 8 $\times$ 8, 9 $\times$ 9.
	\item Data dimension (M) = 25, 36, 49, 64, 81.
	\item Number of atoms (N) = 50, 100, 150, 200.
	\item Sparsity (K) = 3, 5, 7.
	\item Number of iterations ($t_{d}$) = 10.
	\item Pod quantity (P) = 4.
	\item Power iterations ($t_{p}$) = 3
	\item Consensus iterations ($t_{c}$) = 5.
\end{itemize}

\begin{figure}[!htb]
	\centering
	\includegraphics[width=1\linewidth]{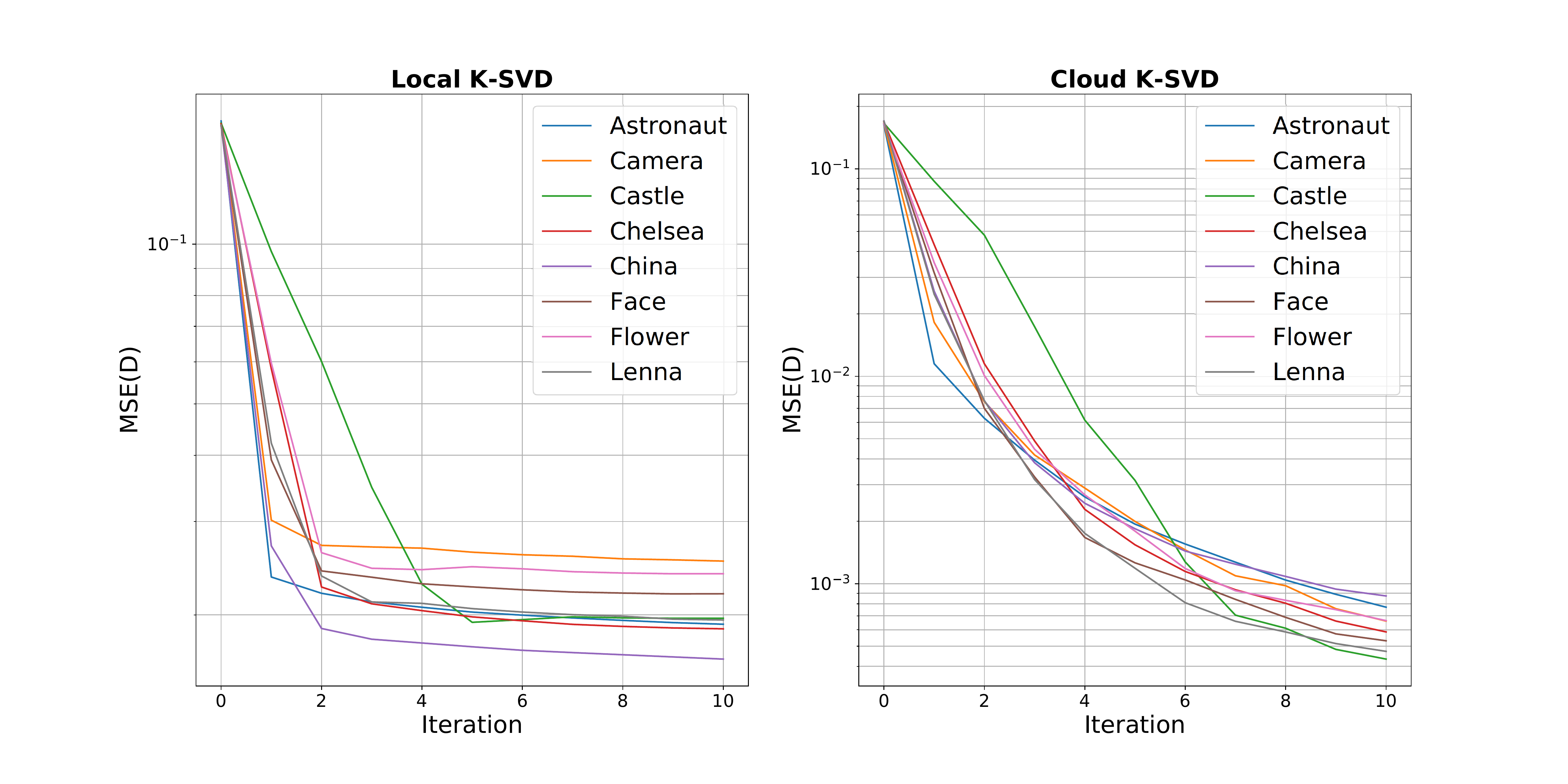}
	\caption{The average MSE between dictionaries of each pod for different noiseless images. $t_{d} = 0, 1, \dots, 10$, $t_{p}=3$, $t_{c}=5$, $M=25$, $N=50$, $K=3$, $P=4$.}
	\label{fig:patch1}
\end{figure}

Figure \ref{fig:patch1} shows the average MSE between dictionaries $\bm{D}_{i} \; \forall i = 1, 2, \dots, P=4$ as a function of $t_{d}$ iterations when consensus is either enabled or disabled. We see a clear drop in MSE between the dictionaries when consensus is enabled (the Cloud K-SVD variant) compared to regular K-SVD with distributed data (the Local K-SVD variant). In other words, the pods learn of each other's data this way thus produce a more mutual dictionary that can be used to represent $\bm{Y}$ better, not only $\bm{Y}_{i}$.

\begin{figure}[!htb]
	\centering
	\includegraphics[width=1\linewidth]{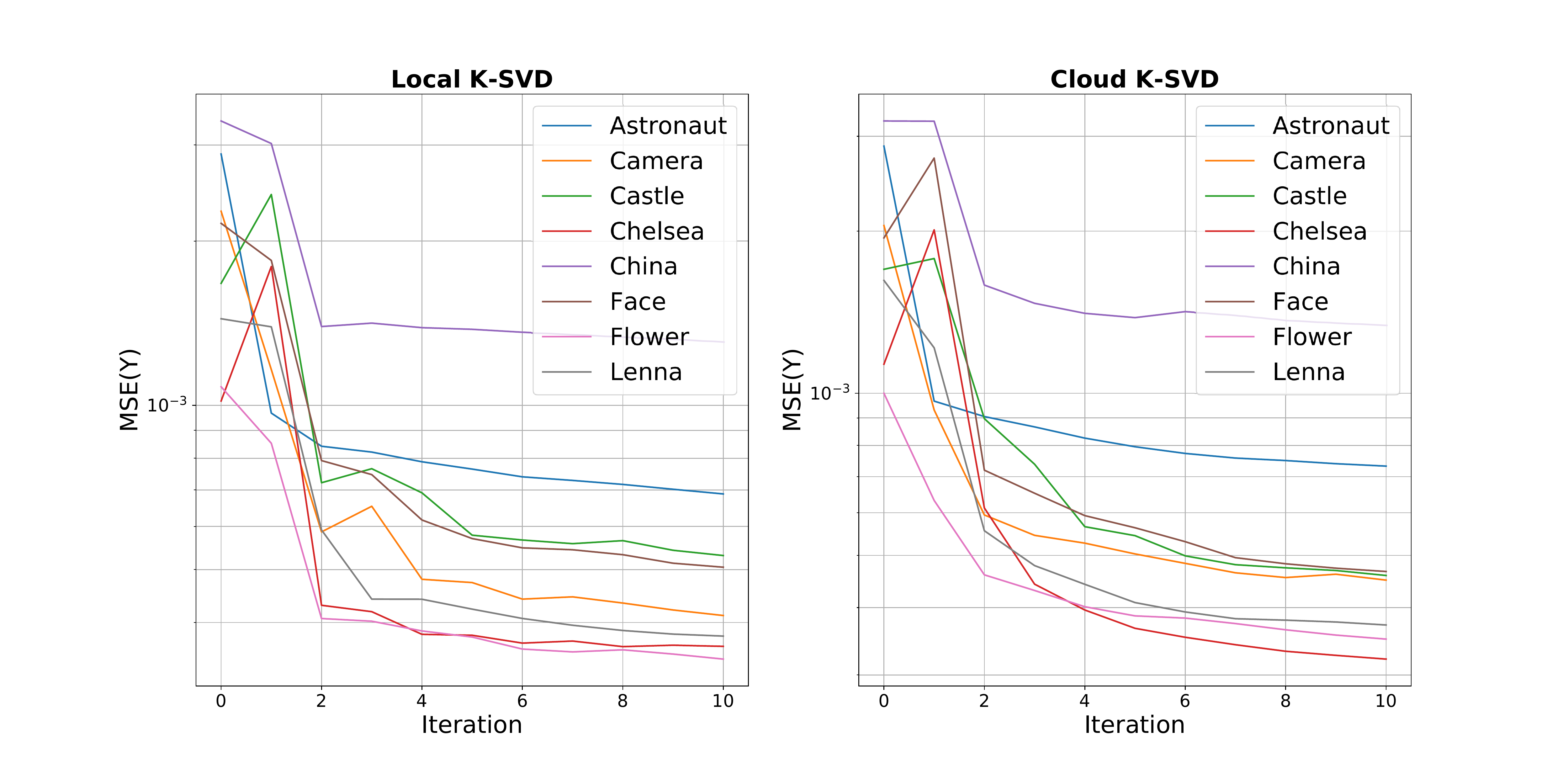}
	\caption{The MSE between the original signal $\bm{Y}$ and the recovered signal $\bm{\hat{Y}}$ for different noiseless images. $t_{d} = 0, 1, \dots, 10$, $t_{p}=3$, $t_{c}=5$, $M=25$, $N=50$, $K=3$, $P=4$.}
	\label{fig:patch5}
\end{figure}

\begin{figure}[!htb]
	\centering
	\includegraphics[width=1\linewidth]{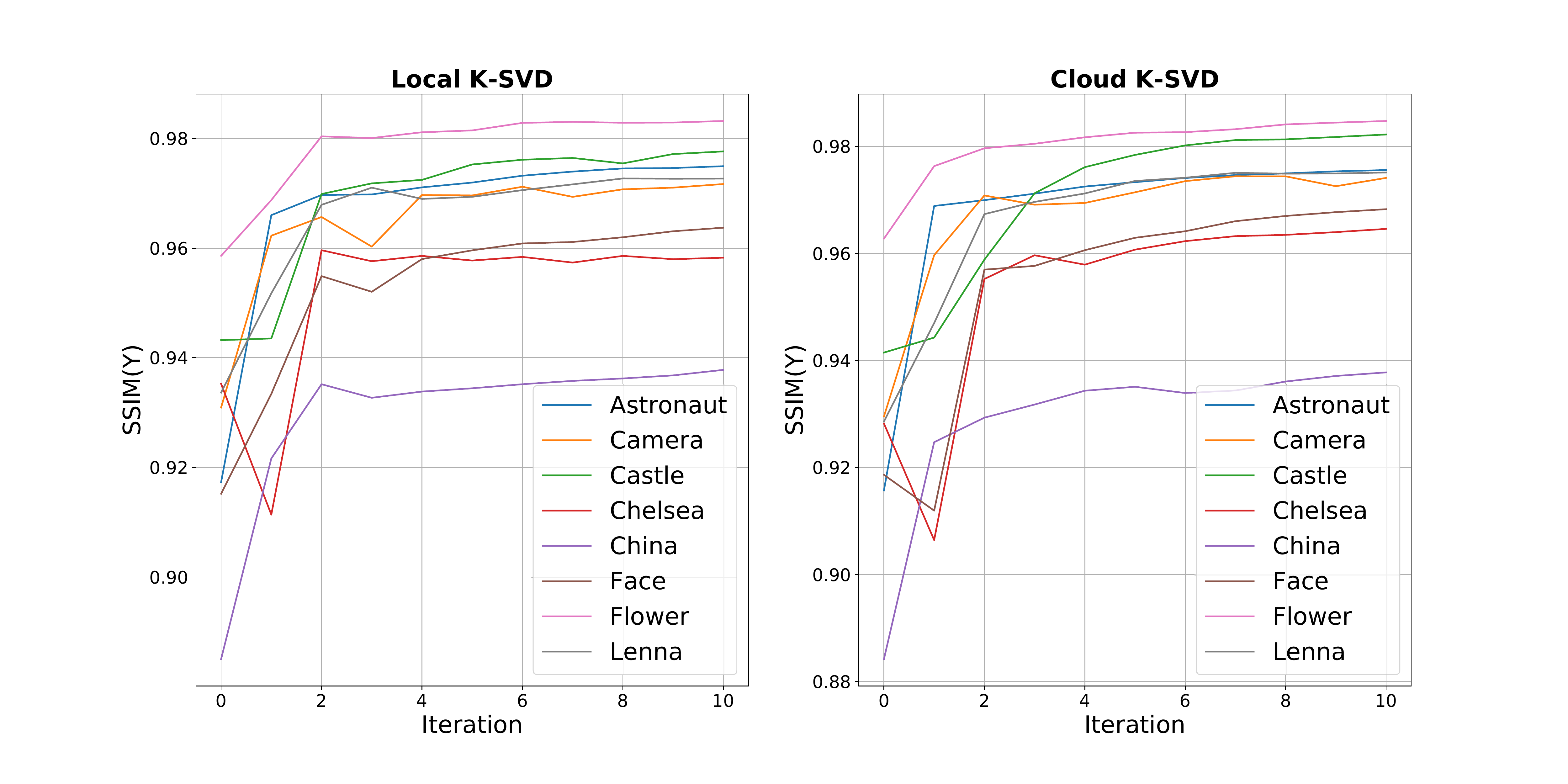}
	\caption{The SSIM between the original signal $\bm{Y}$ and the recovered signal $\bm{\hat{Y}}$ for different noiseless images. $t_{d} = 0, 1, \dots, 10$, $t_{p}=3$, $t_{c}=5$, $M=25$, $N=50$, $K=3$, $P=4$.}
	\label{fig:patch6}
\end{figure}

Figures \ref{fig:patch5} and \ref{fig:patch6} show MSE/SSIM scores when comparing the recovered $\bm{\hat{Y}}$ to $\bm{Y}$ using either Cloud K-SVD or Local K-SVD as a function of iterations $t_{d}$. We observe similar behavior between the two variants, mainly due the fact $\bm{\hat{Y}}_{i}$ is made from a corresponding dictionary $\bm{D}_{i}$ and signal $\bm{X}_{i}$. Here, \textit{China} does worse in terms of MSE/SSIM, whereas we can recover \textit{Flower} a lot better. It seems the algorithm does favor images that contain a lot of repeated contours and few specific structures, but have a lower performance on images with many small and distinctive details.

\begin{figure}[!htb]
	\centering
	\includegraphics[width=1\linewidth,trim={50 90 50 100},clip]{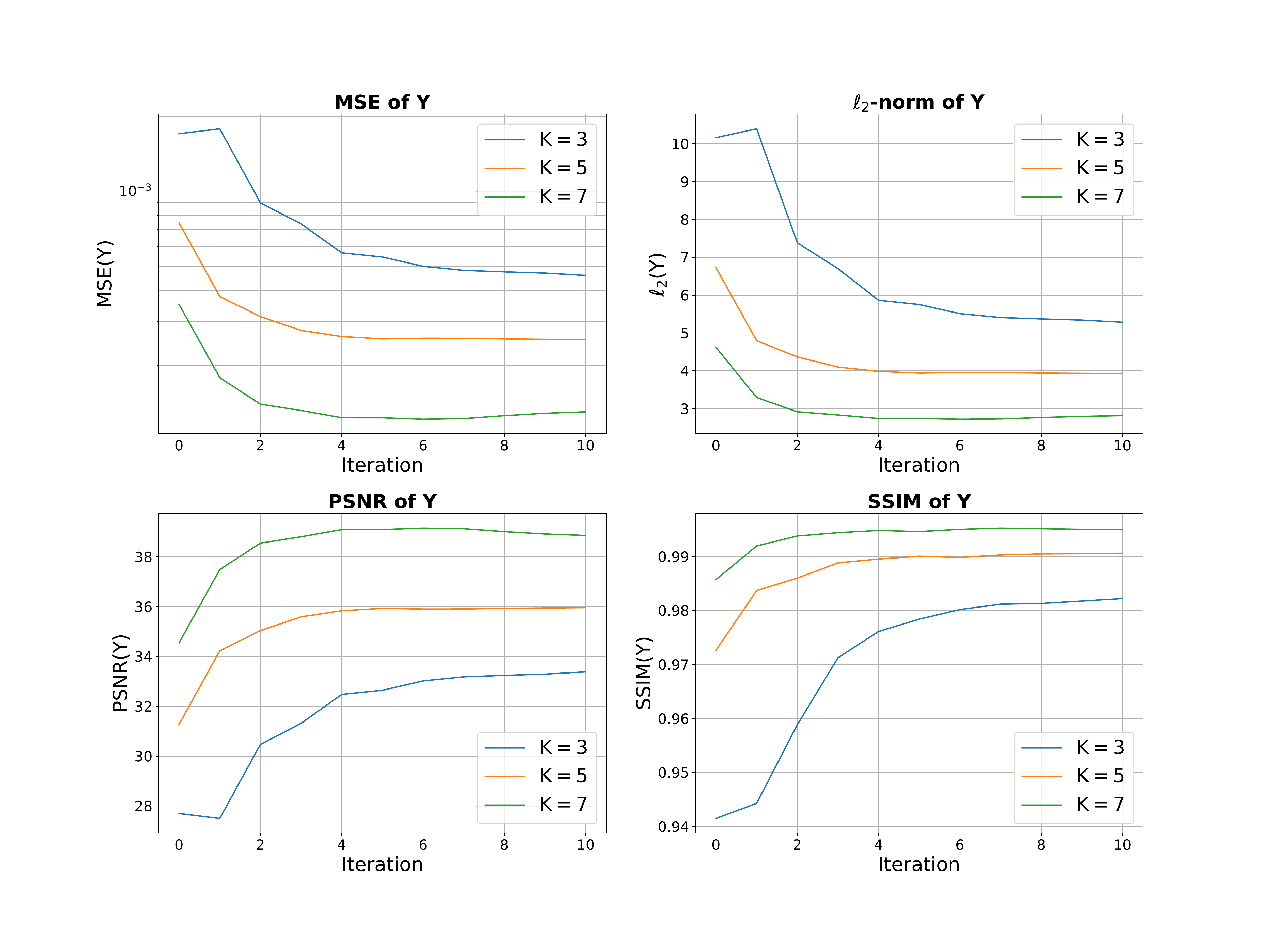}
	\caption{The MSE, $\ell_{2}$-norm of the error, PSNR and SSIM between the original signal $\bm{Y}$ and the recovered signal $\bm{\hat{Y}}$ of the Castle image at different $K$. $t_{d} = 0, 1, \dots, 10$, $t_{p}=3$, $t_{c}=5$, $M=25$, $N=50$, $P=4$.}
	\label{fig:patch11}
\end{figure}

Figure \ref{fig:patch11} shows MSE/$\ell_{2}$-norm/PSNR/SSIM scores when setting the number of nonzero elements $K=3,5,7$ as a function of $t_{d}$ iterations. We notice that a higher $K$ gives lower MSE/$\ell_{2}$-norm and higher PSNR/SSIM scores.

\begin{figure}[!htb]
	\centering
	\includegraphics[width=1\linewidth,trim={50 90 50 100},clip]{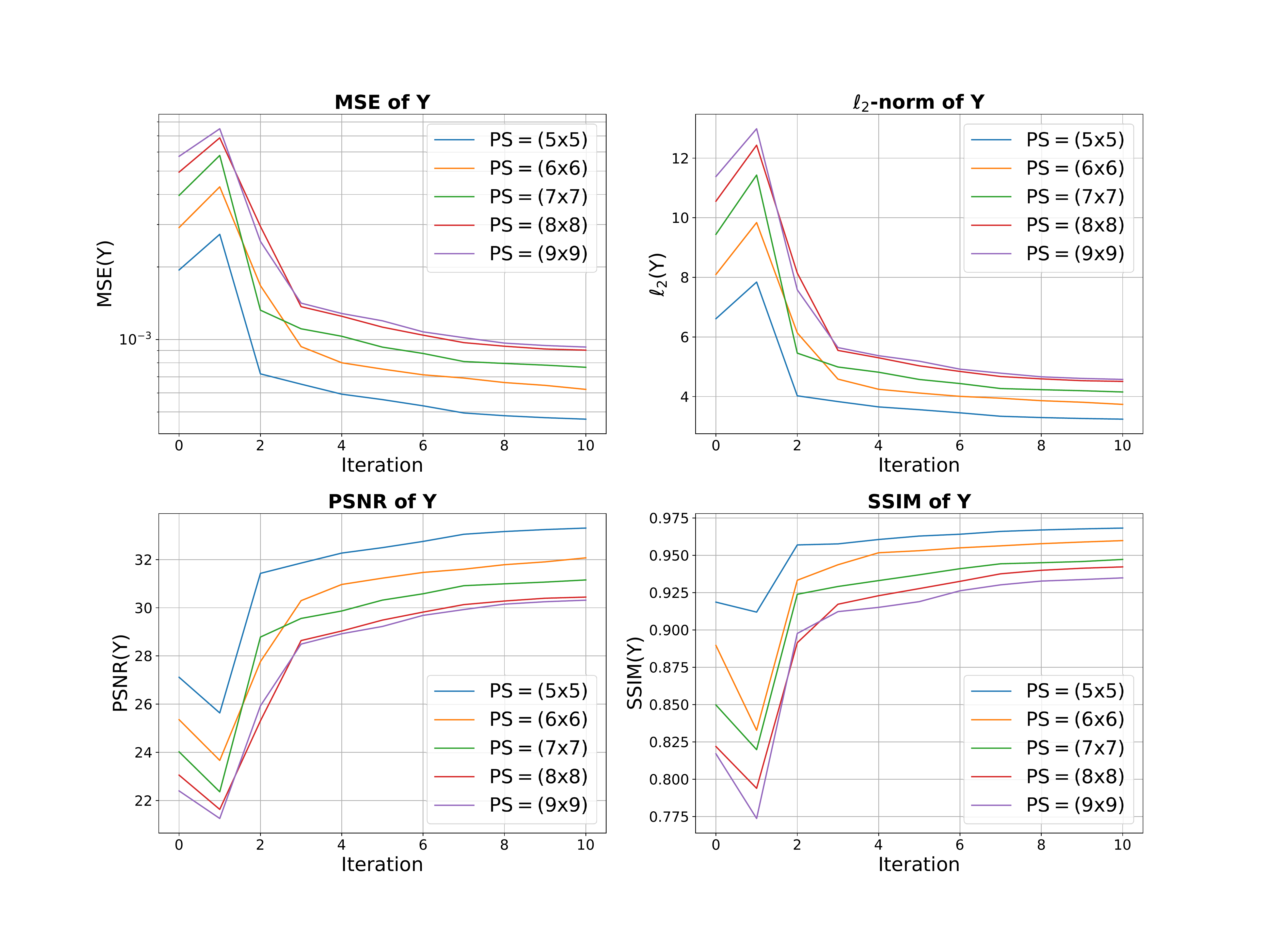}
	\caption{The MSE, $\ell_{2}$-norm of the error, PSNR and SSIM between the original signal $\bm{Y}$ and the recovered signal $\bm{\hat{Y}}$ of the \textit{Face} image at different $M$ and $N$. $t_{d} = 0, 1, \dots, 10$, $t_{p}=3$, $t_{c}=5$, $K=3$, $P=4$.}
	\label{fig:patch12}
\end{figure}

Figure \ref{fig:patch12} shows MSE/$\ell_{2}$-norm/PSNR/SSIM scores as a function of the number of iterations $t_{d}$ when the patch size is increased. Since $PS$ is the data dimension $M$ in number of pixels and $N \gg M$, we increase $N$ as well. We see a lower MSE and $\ell_{2}$-norm between $\bm{Y}$ and $\bm{\hat{Y}}$ when the patch size is lower, for example at $PS=5\times5$/$M=25$, by comparison to a larger patch size, for example $M=49$. The PSNR and SSIM exhibit similar behavior when the patch size goes up. Ideally, we want a low MSE and $\ell_{2}$-norm, but a high PSNR and SSIM.

Tables \ref{tab:lst1}, \ref{tab:lst2} and \ref{tab:lst3} show average execution times for four pods tracked for the OMP and K-SVD step for values of $K=3, 5, 7$, $M=25, 36, 49, 64, 81$ and $\alpha$=1, 2 in Local K-SVD and Cloud K-SVD runs, respectively. The OMP step seems to be impacted significantly by a higher $K$, as expected, while the K-SVD step is doing acceptable with $K$ set high. OMP is not slowed down by using larger patch dimensionality $M$, but K-SVD is. This is because generally $N \gg M$, usually twice its size, so the dictionary size $N$ gets bigger as $M$ go up. Because K-SVD updates the dictionary per atom, it naturally takes longer to update a large dictionary than a small one.

\normalsize
\begin{table*}[!htb]
	\begin{minipage}{1\linewidth}
		\caption{Average execution times of the OMP and K-SVD step for Local K-SVD ($t_{p}=0, t_{c}=0$) and Cloud K-SVD ($t_{p}=3, t_{c}=5$) using noiseless patches. $K=3$, $M=25$, $N=50$. \textit{Castle} and \textit{Face} have $\alpha=1$, the rest have $\alpha=2$}
		\label{tab:lst1}
		\begin{tabular}{lllllllll} \toprule
			\normalsize
			Variant/step & {\textbf{Astronaut}} & {\textbf{Camera}} & {\textbf{Castle}} & {\textbf{Chelsea}} & {\textbf{China}} & {\textbf{Face}} & {\textbf{Flower}} & {\textbf{Lenna}} \\ \midrule
			Local/OMP      & 28s	& 29.4s	& 27.3s	& 15.1s	& 30.6s	& 9.8s	& 30.7s	& 27.6s   \\
			Local/K-SVD    & 7.5s	& 7.1s	& 6.7s	& 3.3s	& 8s	& 2.5s	& 7.6s	& 6.8s	  \\
			Cloud/OMP      & 34.2s	& 32.6s	& 31.6s	& 18.5s	& 34.2s	& 12.5s	& 35.6s	& 31.2s	  \\
			Cloud/K-SVD    & 42.3s	& 43.1s	& 43.5s	& 33.9s	& 44.9s	& 31.9s	& 46.4s	& 42.8s   \\ \bottomrule
		\end{tabular}
	\end{minipage}
\end{table*}

\begin{table}[!htb]
	\begin{minipage}{1\linewidth}
		\caption{Average execution times of the OMP and K-SVD step for Local K-SVD ($t_{p}=0, t_{c}=0$) and Cloud K-SVD variants ($t_{p}=3, t_{c}=5$) using \textit{Castle} with different $K$ values. $M=25$, $N=50$, $\alpha=1$.}
		\label{tab:lst2}
		\begin{tabular}{llll} \toprule
			Variant/step & \pbox{20cm}{\textbf{Castle} \\ $K=3$} & \pbox{20cm}{\textbf{Castle} \\ $K=5$} & \pbox{20cm}{\textbf{Castle} \\ $K=7$} \\ \midrule
			Local/OMP & 27.3s	& 45.2s	& 64.4s		\\
			Local/K-SVD & 6.7s	& 8.7s	& 11.1s		\\
			Cloud/OMP & 31.6s	& 51.9s	& 74.2s		\\
			Cloud/K-SVD & 43.5s	& 45.4s	& 43.1s		\\ \bottomrule
		\end{tabular}
	\end{minipage}
\end{table}

\begin{table*}[!htb]
	\begin{minipage}{1\linewidth}
		\centering
		\caption{Average execution times of the OMP and K-SVD step for Local K-SVD ($t_{p}=0, t_{c}=0$) and Cloud K-SVD ($t_{p}=3, t_{c}=5$) variants using \textit{Face} with different $M$ and $N$ values. $K=3$. $\alpha=1$.}
		\label{tab:lst3}
		\begin{tabular}{llllll} \toprule
			Variant/iteration & \pbox{20cm}{\textbf{Face} \\ $M=25$ \\ $N=50$} & \pbox{20cm}{\textbf{Face} \\ $M=36$ \\ $N=100$} & \pbox{20cm}{\textbf{Face} \\ $M=49$ \\ $N=100$} & \pbox{20cm}{\textbf{Face} \\ $M=64$ \\ $N=150$} & \pbox{20cm}{\textbf{Face} \\ $M=81$ \\ $N=200$} \\ \midrule
		Local/OMP	& 9.8s	& 10s		& 10.1s	& 10.6s	& 10.9s	\\
		Local/K-SVD& 2.5s	& 4.8s	& 6.6s	& 12.9s	& 17.7s	\\
		Cloud/OMP	& 12.5s	& 14.3s	& 13.3s	& 16.3s	& 15.9s	\\
		Cloud/K-SVD& 31.9s	& 63.4s	& 71.5s	& 120.5s	& 157.9s	\\ \bottomrule
		\end{tabular}
	\end{minipage}
\end{table*}

\subsection{Experiment: Denoising}

\begin{figure}[!htb]
	\centering
	\includegraphics[width=1\linewidth,trim={170 195 140 145},clip]{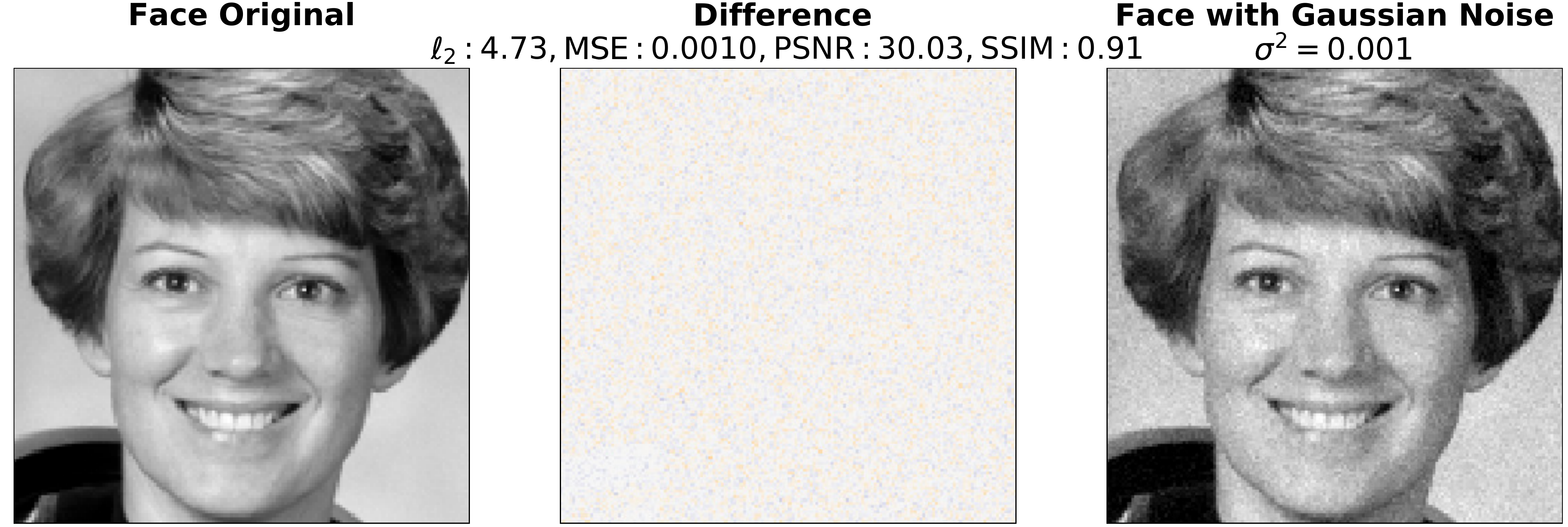}
	\caption{Left: Original. Middle: Difference. Right: Noisy ($\mu = 0, \sigma^{2} = 0.001$).}
	\label{fig:denoise1}
\end{figure}

\begin{figure}[!htb]
	\centering
	\includegraphics[width=1\linewidth,trim={170 195 140 145},clip]{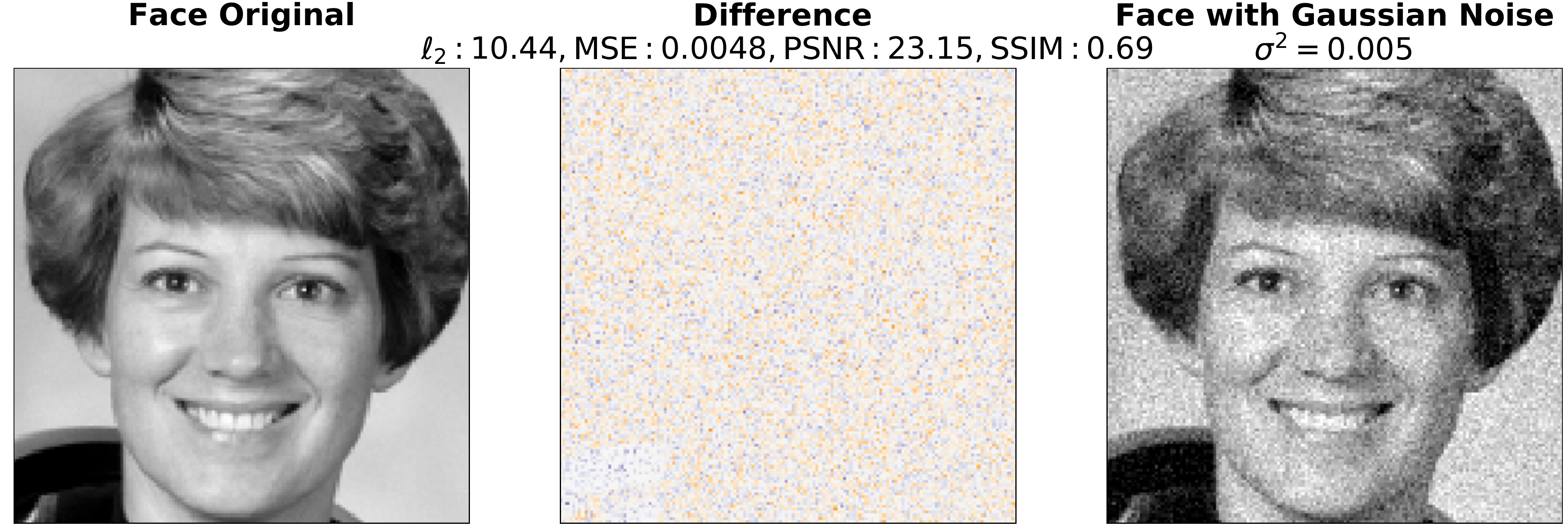}
	\caption{Left: Original. Middle: Difference. Right: Noisy ($\mu = 0, \sigma^{2} = 0.005$).}
	\label{fig:denoise2}
\end{figure}

\begin{figure}[!htb]
	\centering
	\includegraphics[width=1\linewidth,trim={170 195 140 145},clip]{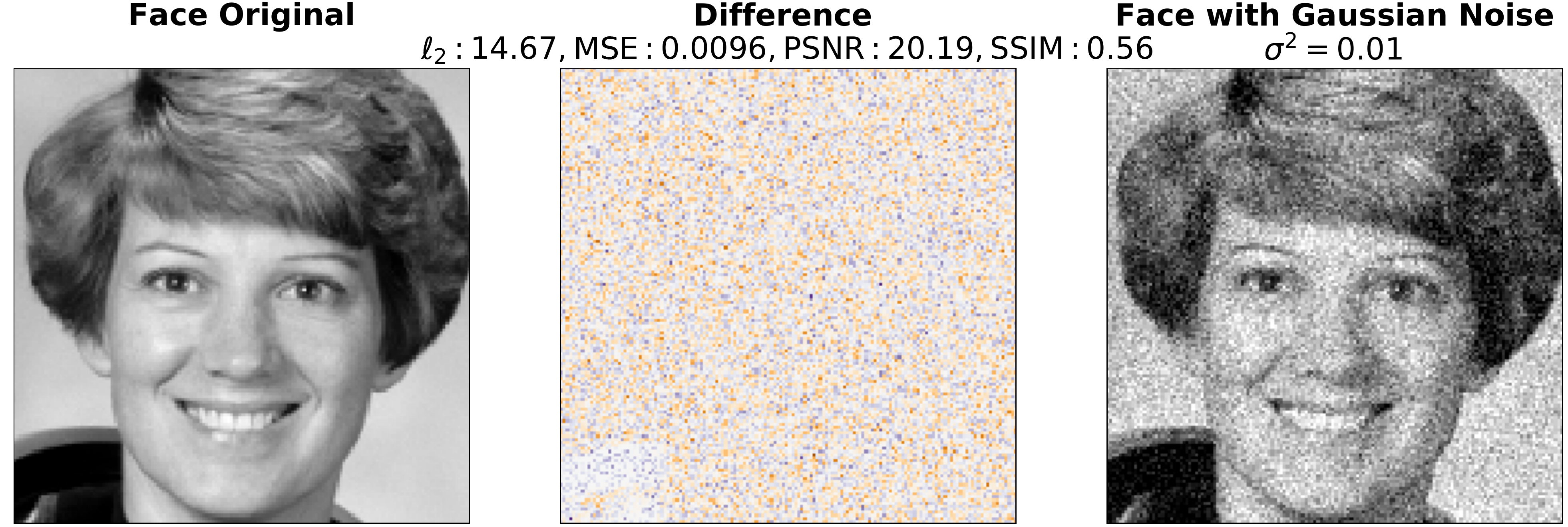}
	\caption{Left: Original. Middle: Difference. Right: Noisy ($\mu = 0, \sigma^{2} = 0.01$).}
	\label{fig:denoise3}
\end{figure}

We now add a layer of Gaussian distributed noise (AWGN) by mean $\mu = 0$ and variance $\sigma^{2} = 0.001, 0.005, 0.01$ to all training images and then extract $Q$ noisy patches. Figures \ref{fig:denoise1}, \ref{fig:denoise2} and \ref{fig:denoise3} show the \textit{Face} image in a clean and noisy version.

\begin{figure}[!htb]
	\centering
	\includegraphics[width=1\linewidth,trim={50 130 50 120},clip]{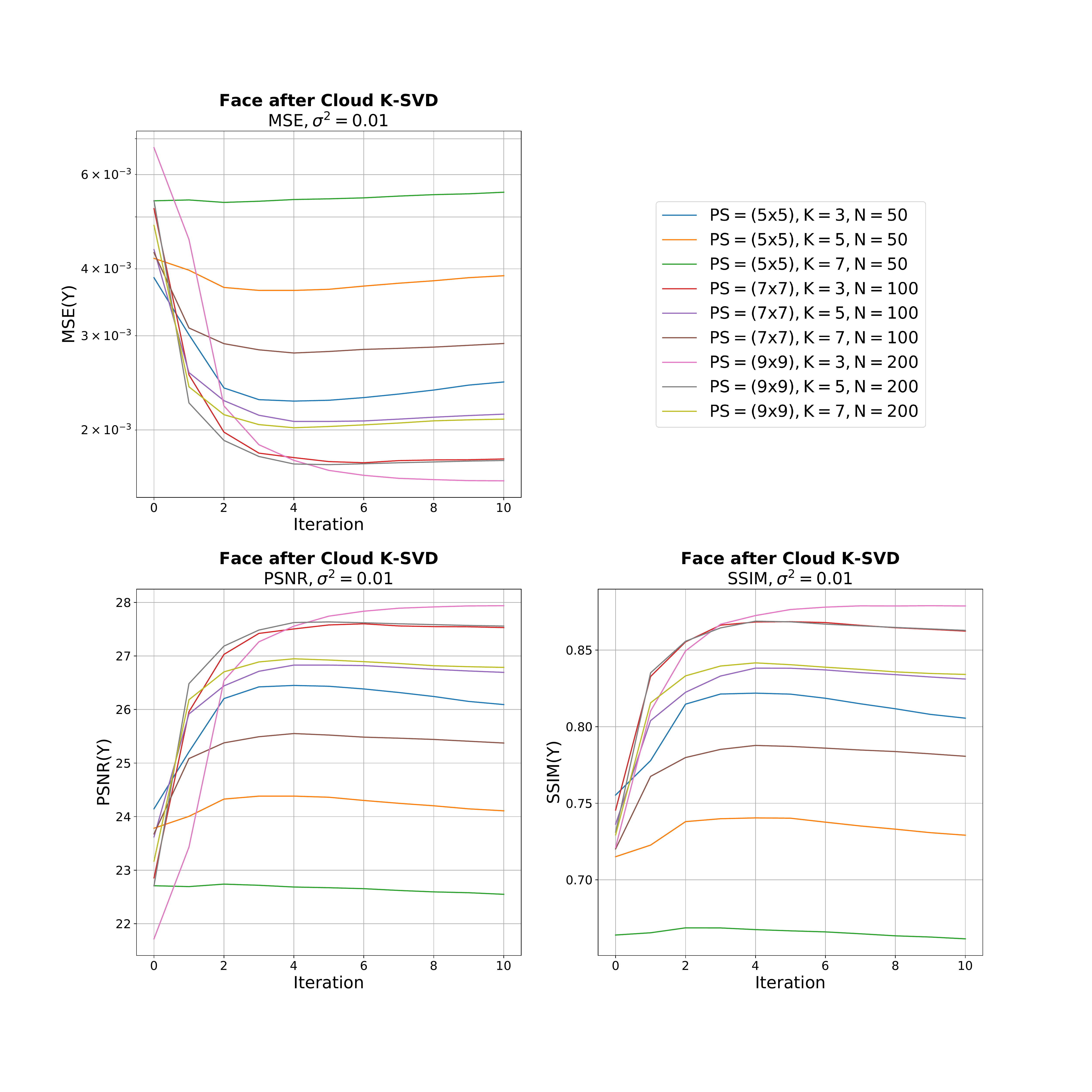}
	\caption{The MSE, PSNR and SSIM between the original signal $\bm{Y}$ and the recovered signal $\bm{\hat{Y}}$ of the \textit{Face} image with AWGN ($\sigma^{2}=0.01$) at different $M$, $N$ and $K$. $t_{d} = 0, 1, \dots, 10$, $t_{p}=3$, $t_{c}=5$, $P=4$.}
	\label{fig:denoise4}
\end{figure}

\begin{figure}[!htb]
	\centering
	\includegraphics[width=1\linewidth,trim={50 150 50 100},clip]{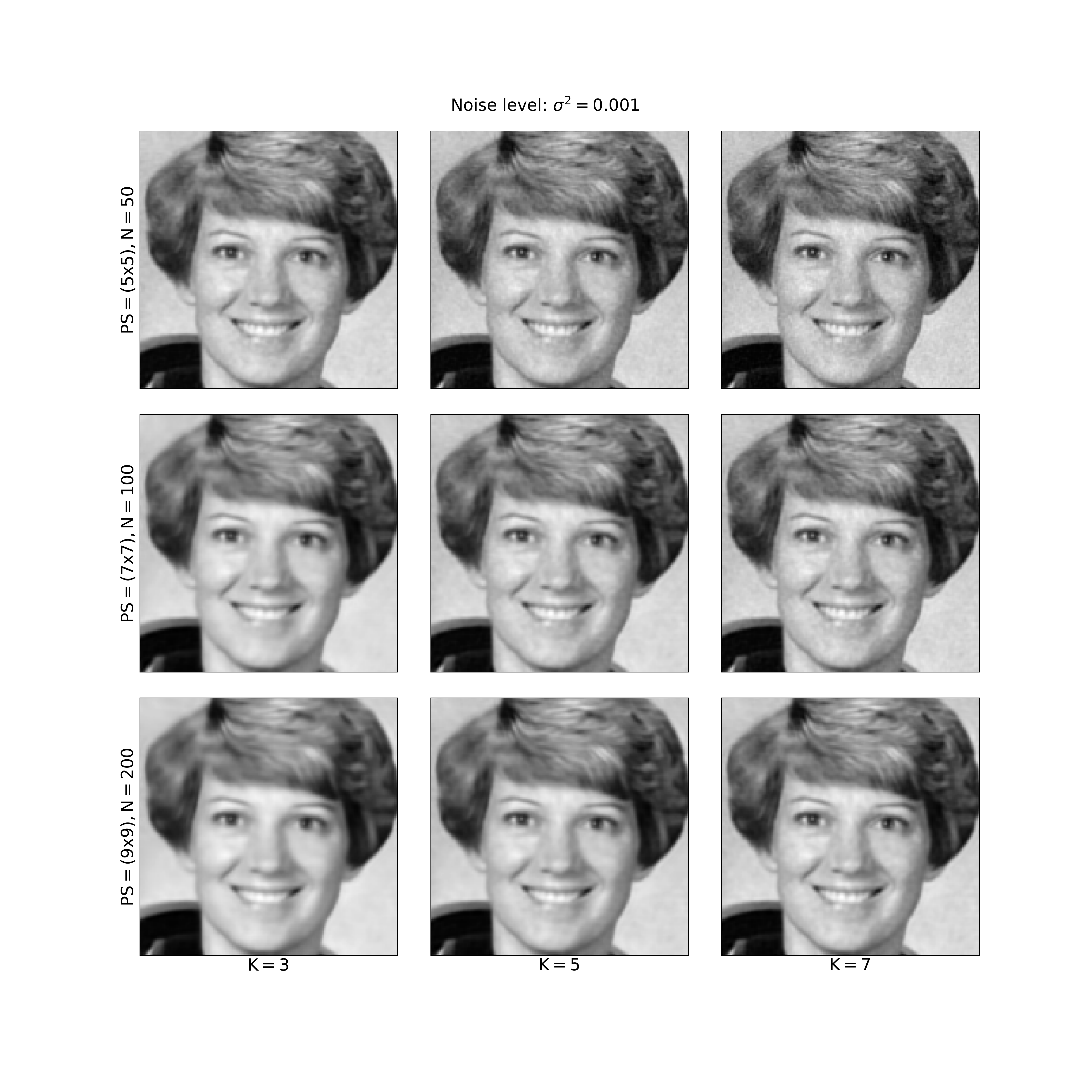}
	\caption{A collage of recovered \textit{Face} images using Cloud K-SVD at varying $M$, $N$ and $K$. $\sigma^{2} = 0.001$, $t_{d}=10$, $t_{p}=3$, $t_{c}=5$, $P=4$.}
	\label{fig:denoise5}
\end{figure}

Figure \ref{fig:denoise4} shows MSE/PSNR/SSIM scores for $\bm{\hat{Y}}$ as a function of $t_{d}$ iterations compared to the original noiseless $\bm{Y}$ after Cloud K-SVD has recovered the \textit{Face} image from noisy patches. Figure \ref{fig:denoise5} is a collage of recovered images. Here, the variance is set to $\sigma^{2} = 0.01$ with various configurations of $PS$, $K$ and $N$. We see a break-even around the forth iteration for most cases and some configurations even tend to do worse as the number of iterations go up. Because the AWGN level is quite high, at some iteration the algorithm starts recovering the noise because it has only been exposed to these noisy measurements. For comparison with \cite{Ghosh2017}, our noise variance of $\sigma^{2} = 0.01$ would mean a $\sigma=25.5$ using \cite{Ghosh2017}'s notation, since $\sqrt{0.01}*255=25.5$. The best PSNR/SSIM scores we obtain are approximately \textbf{27.98/88.05} with $PS=9\times9, K=3, N=200$ and $\sigma=25.5$ on the \textit{Face} image (using \cite{Ghosh2017}'s notation). \cite{Ghosh2017} obtain PSNR/SSIM scores of \textbf{29.21/85.64} for $\sigma=20$ and \textbf{27.67/79.93} for $\sigma=30$ when denoising the \textit{Hill} image \cite{BM3D}. Hence we obtain comparable PSNR/SSIM scores to \cite{Ghosh2017}, however note that the patch size is different (we use $PS=9\times9$, they use $PS=3\times3$).

\begin{figure}[!htb]
	\centering
	\includegraphics[width=1\linewidth,trim={50 130 50 100},clip]{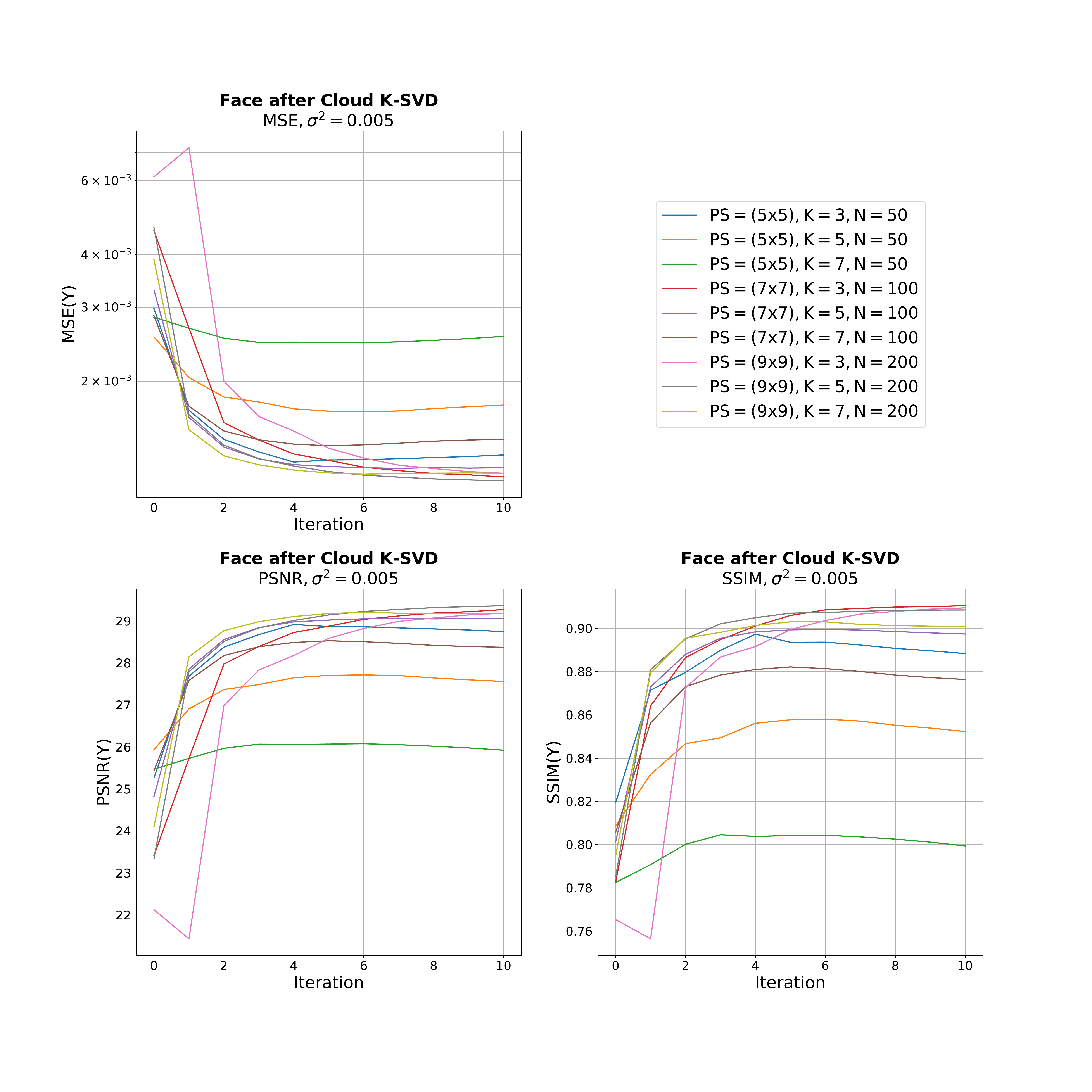}
	\caption{The MSE, PSNR and SSIM between the original signal $\bm{Y}$ and the recovered signal $\bm{\hat{Y}}$ of the \textit{Face} image with AWGN ($\sigma^{2}=0.005$) at different $M$, $N$ and $K$. $t_{d} = 0, 1, \dots, 10$, $t_{p}=3$, $t_{c}=5$, $P=4$.}
	\label{fig:denoise6}
\end{figure}

\begin{figure}[!htb]
	\centering
	\includegraphics[width=1\linewidth,trim={50 150 50 100},clip]{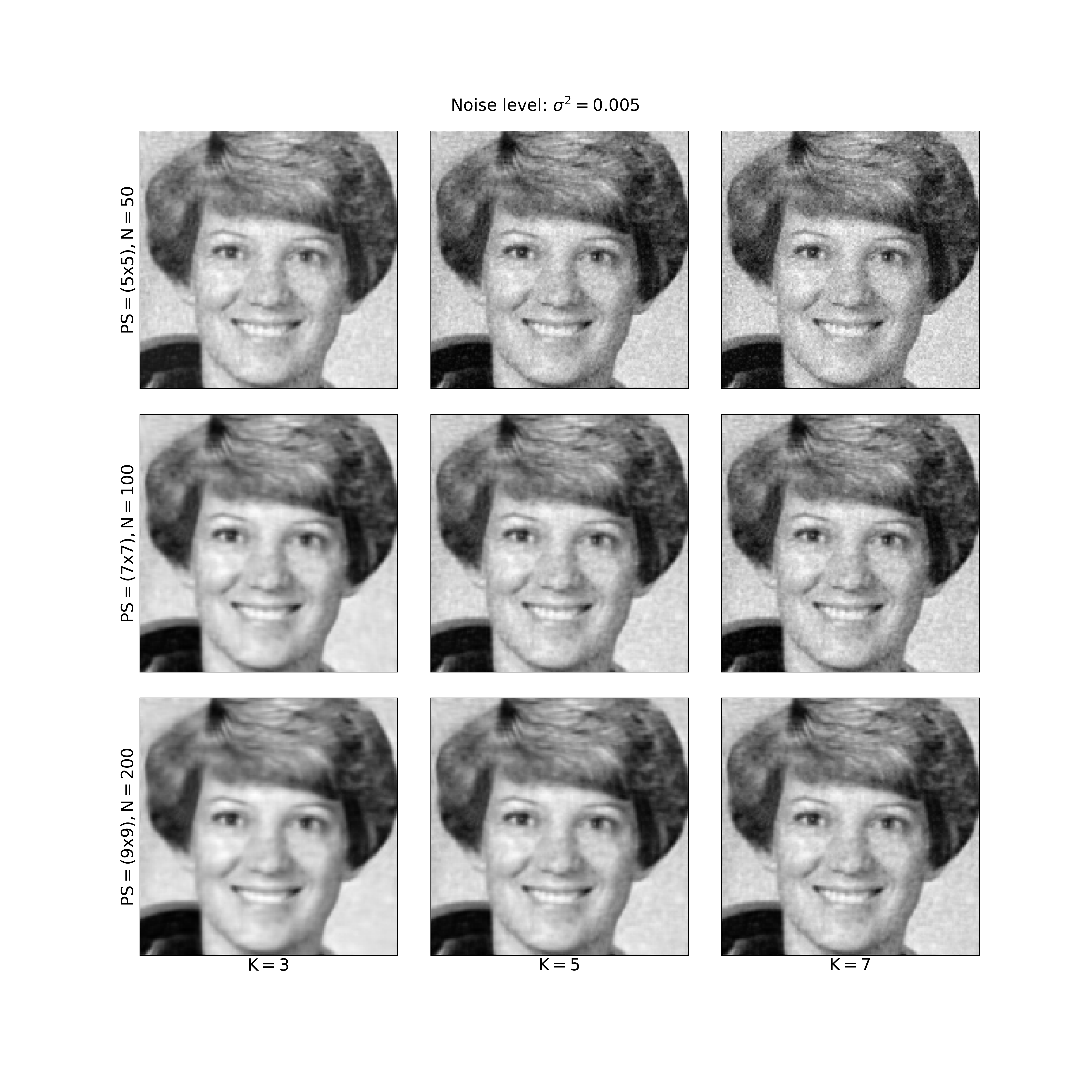}
	\caption{A collage of recovered \textit{Face} images using Cloud K-SVD at varying $M$, $N$ and $K$. $\sigma^{2} = 0.005$, $t_{d}=10$, $t_{p}=3$, $t_{c}=5$, $P=4$.}
	\label{fig:denoise7}
\end{figure}

Figure \ref{fig:denoise6} shows MSE/PSNR/SSIM scores with $\sigma^{2} = 0.005$. Clearly, a low patch size and a high sparsity (the green line) struggles to reduce AWGN, whilst a low sparsity configuration does a better job. By looking at the results, an acceptable configuration is a patch size of $7 \times 7$ thus $M=49$, an atom count of $100$ and a sparsity of $K=3$. Figure \ref{fig:denoise7} is a collage of recovered images. For comparison, $\sigma^{2} = 0.005$ equals $\sigma=18$ in \cite{Ghosh2017}'s notation. The best PSNR/SSIM scores we obtain are approximately \textbf{29.50/91.05} with $PS=9x9, K=5, N=200$ and $\sigma=18$. \cite{Ghosh2017} obtain \textbf{29.21/85.64} with $\sigma=20$. Again, our image of reference and patch size is different (\textit{Face} vs. \textit{Hill} \cite{BM3D}, $PS=9\times9$ vs. $3\times3$).

\begin{figure}[!htb]
	\centering
	\includegraphics[width=1\linewidth,trim={50 130 50 100},clip]{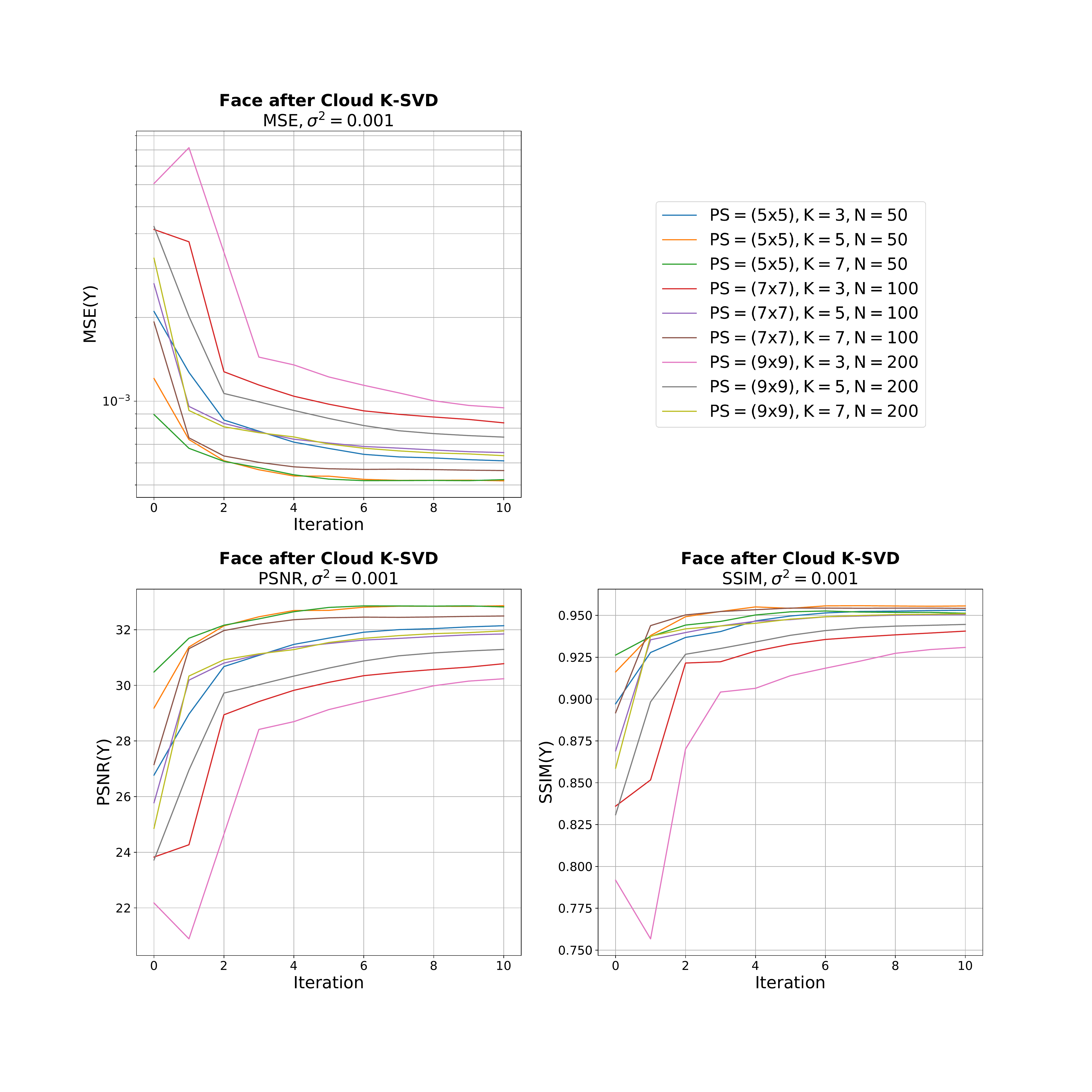}
	\caption{The MSE, PSNR and SSIM between the original signal $\bm{Y}$ and the recovered signal $\bm{\hat{Y}}$ of the \textit{Face} image with AWGN ($\sigma^{2}=0.001$) at different $M$, $N$ and $K$. $t_{d} = 0, 1, \dots, 10$, $t_{p}=3$, $t_{c}=5$, $P=4$.}
	\label{fig:denoise8}
\end{figure}

\begin{figure}[!htb]
	\centering
	\includegraphics[width=1\linewidth,trim={50 150 50 100},clip]{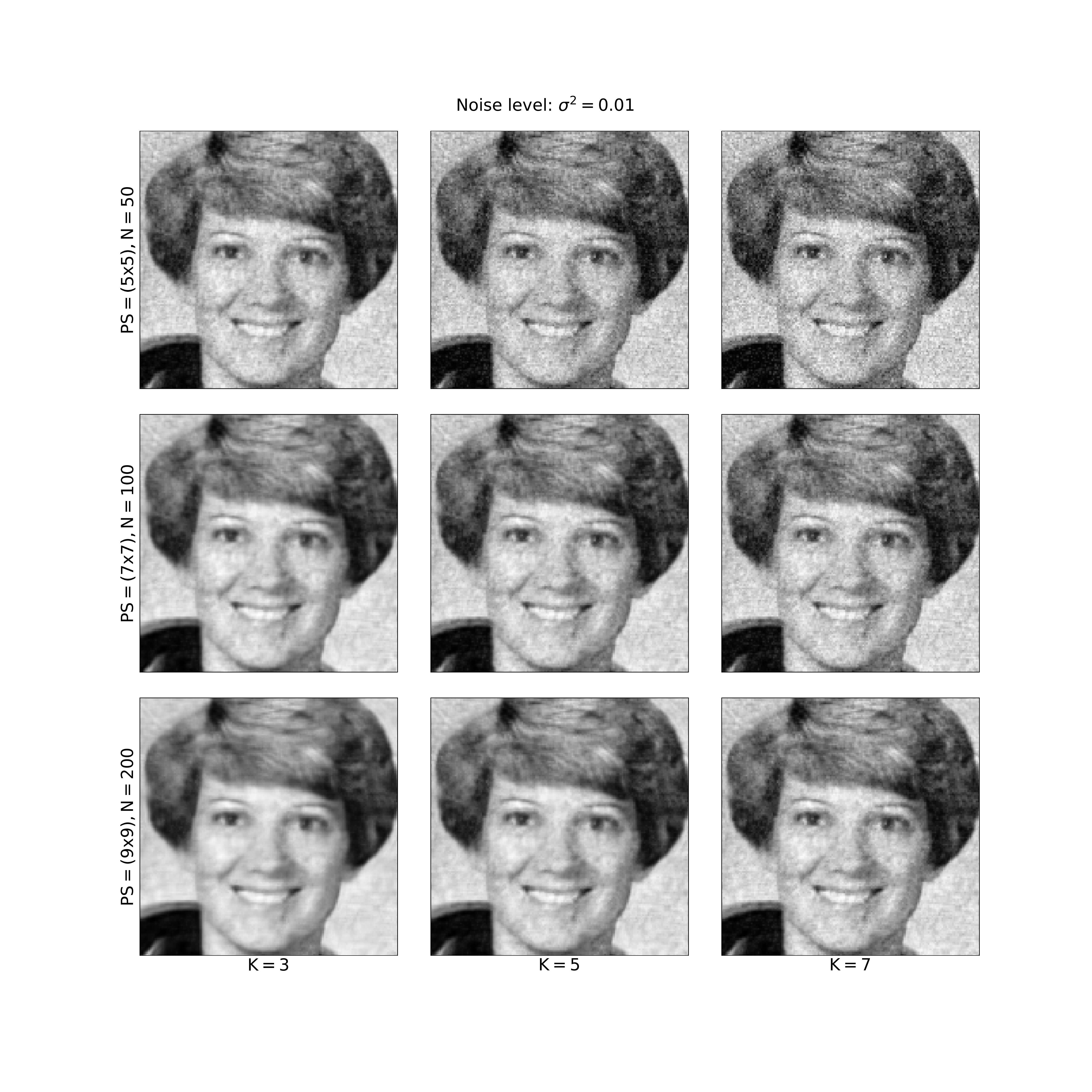}
	\caption{A collage of recovered \textit{Face} images using Cloud K-SVD at varying $M$, $N$ and $K$. $\sigma^{2} = 0.01$, $t_{d}=10$, $t_{p}=3$, $t_{c}=5$, $P=4$.}
	\label{fig:denoise9}
\end{figure}

Figure \ref{fig:denoise8} shows MSE/PSNR/SSIM scores with $\sigma^{2} = 0.001$. A high pixel size $PS$ with a low sparsity $K$ setting is no longer ideal when the AWGN variance is low. When $K$ is low and $PS$ is high (the pink line), Cloud K-SVD struggles to properly recover the image. Cloud K-SVD performs better with high sparsity in case there is little to no noise in the input signal. Figure \ref{fig:denoise9} is a collage of recovered images. For comparison, $\sigma^{2} = 0.001$ equals $\sigma=8$ in \cite{Ghosh2017}'s notation. The best PSNR/SSIM scores we obtain are approximately \textbf{32.90/95.10} with $PS=5x5, K=7, N=50$ and $\sigma=8$. \cite{Ghosh2017} obtain \textbf{32.15/94.01} with $\sigma=10$. Again, our image of reference and patch size is different (\textit{Face} vs. \textit{Hill} \cite{BM3D}, $PS=5\times5$ vs. $3\times3$).

Table \ref{tab:lst4} shows average execution times for four pods tracked for the OMP and K-SVD step.

\begin{table*}[!htb]
	\begin{minipage}{1\linewidth}
		\centering
		\caption{Average execution times of the OMP and K-SVD step for Cloud K-SVD ($t_{p}=3, t_{c}=5$) when recovering \textit{Face} from noisy patches with different sparsity $K$, data size $M$ and AWGN level $\sigma^{2}$. We set $\alpha=1$.}
		\label{tab:lst4}
		\begin{tabular}{llllllllll} \toprule
			Step/$\sigma^{2}$
			 & \pbox{20cm}{\textbf{Face} \\ $K=3$ \\ $M=25$}
 			 & \pbox{20cm}{\textbf{Face} \\ $K=3$ \\ $M=49$}
 			 & \pbox{20cm}{\textbf{Face} \\ $K=3$ \\ $M=81$}
			 & \pbox{20cm}{\textbf{Face} \\ $K=5$ \\ $M=25$}
			 & \pbox{20cm}{\textbf{Face} \\ $K=5$ \\ $M=49$}
			 & \pbox{20cm}{\textbf{Face} \\ $K=5$ \\ $M=81$}
			 & \pbox{20cm}{\textbf{Face} \\ $K=7$ \\ $M=25$}
			 & \pbox{20cm}{\textbf{Face} \\ $K=7$ \\ $M=49$}
			 & \pbox{20cm}{\textbf{Face} \\ $K=7$ \\ $M=81$} \\ \midrule
			OMP/$0.001$	& 11.9s	& 12.4	& 14.1	& 20s & 20.1s  & 22s    & 28s    & 28.3s    & 33.2s	 \\
			OMP/$0.005$	& 12.5	& 12.8s	& 14.9s	& 19s & 21.5s  & 23.7s  & 27.4s  & 29.8s    &  31.7s	\\
			OMP/$0.01$	& 12.3  & 12.2s & 14.7s	& 20.5s  & 22.4s  & 22.2s & 28.5s & 28.8s  & 32.5s	\\
			K-SVD/$0.001$ & 31.5s & 72.1 & 178.1s & 33s	& 69.5s  & 178.9s    & 33.7s     & 70.3s     & 179.6s	\\
			K-SVD/$0.005$ & 32.1s & 68.7s & 179s  & 32.5s & 67.8s   & 175.9s   & 33.3s & 67.5s  & 178.5s 	\\
			K-SVD/$0.01$  & 31.8s & 70s	  & 160.3s  & 32.7s  & 70.9s  & 159.5s & 33.3s & 73.9s  & 158.4s 	\\ \bottomrule
		\end{tabular}
	\end{minipage}
\end{table*}

\section{Conclusion}
\label{sec:conclusion}

In order to successfully perform AWGN denoising, the parameters of Cloud K-SVD need be adjusted accordingly. When the noise is severe, a lot of detail have to be removed, however when it is less severe, less detail need be removed. Details in images can be recovered better by decreasing data size and atom count, $M$ and $N$, and increasing the number of nonzero elements, $K$. Cloud K-SVD was originally made for image classification, but we have made a novel application of the algorithm and some practical improvements, e.g. used Simultaneous Orthogonal Matching Pursuit (SOMP) and adjusted the algorithm to account for node failures. We show, that our modified version of Cloud K-SVD can successfully conduct distributed AWGN denoising of images and simultaneously learn a global dictionary. We compared our results to the Pruned Non-Local Means (PNLM) algortihm presented in \cite{Ghosh2017}, which supports our claim that the accuracy of recovery in Cloud K-SVD is on par with SOTA in the field.

\section*{Competing Interests}

The authors declare that they have no known competing financial interests or personal relationships that could have appeared to influence the work reported in this paper.

\interlinepenalty=10000
\bibliography{sn-bibliography}

\end{document}